\documentclass[aip,amsmath,amssymb,reprint,floatfix]{revtex4-1}

\usepackage{graphicx}
\usepackage{dcolumn}
\usepackage{bm}
\usepackage{siunitx}

\newcounter{Cequ}
\newenvironment{CEquation}
  {\stepcounter{Cequ}%
    \addtocounter{equation}{-1}%
    \renewcommand\theequation{S\arabic{Cequ}}\equation}
  {\endequation}

\begin{document}

\preprint{AIP/123-QED}
\title[Scaling waveguide-integrated SNSPD solutions to large numbers of independent optical channels]{Scaling waveguide-integrated superconducting nanowire single-photon detector solutions to large numbers of independent optical channels}

\author{Matthias Häußler}
\email{matthias.haeussler@uni-muenster.de}
\author{Robin Terhaar}
\altaffiliation[]{Contributed equally to this work}
\author{Martin A. Wolff}
\author{Helge Gehring}
\affiliation{Institute of Physics, University of Münster, Heisenbergstraße 11, 48149 Münster, Germany}
\affiliation{Center for NanoTechnology (CeNTech), Heisenbergstraße 11, 48149 Münster, Germany}
\affiliation{Center for Soft Nanoscience (SoN), Busso-Peus Straße 10, 48149 Münster, Germany}
\author{Fabian Beutel}
\author{Wladick Hartmann}
\author{Nicolai Walter}
\affiliation{Institute of Physics, University of Münster, Heisenbergstraße 11, 48149 Münster, Germany}
\affiliation{PixelPhotonics GmbH, Heisenbergstraße 11, 48149 Münster, Germany}
\author{Max Tillmann}
\author{Mahdi Ahangarianabhari}
\author{Michael Wahl}
\author{Tino Röhlicke}
\author{Hans-Jürgen Rahn}
\affiliation{PicoQuant GmbH, Rudower Chaussee 29, 12489 Berlin, Germany}
\author{Wolfram H.P. Pernice}
\affiliation{Institute of Physics, University of Münster, Heisenbergstraße 11, 48149 Münster, Germany}
\affiliation{Center for NanoTechnology (CeNTech), Heisenbergstraße 11, 48149 Münster, Germany}
\affiliation{Center for Soft Nanoscience (SoN), Busso-Peus Straße 10, 48149 Münster, Germany}
\affiliation{Kirchhoff-Institute for Physics, Heidelberg University, Im Neuenheimer Feld 227, 69120 Heidelberg, Germany}
\author{Carsten Schuck}
\email{carsten.schuck@uni-muenster.de}
\affiliation{Institute of Physics, University of Münster, Heisenbergstraße 11, 48149 Münster, Germany}
\affiliation{Center for NanoTechnology (CeNTech), Heisenbergstraße 11, 48149 Münster, Germany}
\affiliation{Center for Soft Nanoscience (SoN), Busso-Peus Straße 10, 48149 Münster, Germany}

\begin{abstract}
Superconducting nanowire single-photon detectors are an enabling technology for modern quantum information science and are gaining attractiveness for the most demanding photon counting tasks in other fields. Embedding such detectors in photonic integrated circuits enables additional counting capabilities through nanophotonic functionalization. Here we show how a scalable number of waveguide-integrated superconducting nanowire single-photon detectors can be interfaced with independent fiber optic channels on the same chip. Our plug-and-play detector package is hosted inside a compact and portable closed-cycle cryostat providing cryogenic signal amplification for up to 64 channels. We demonstrate state-of-the-art photon counting performance with up to 60\,\% system detection efficiency and down to 
26.0\,ps timing accuracy for individually addressable detectors. Our multi-channel single photon receiver offers exciting measurement capabilities for future quantum communication, remote sensing and imaging applications. 
\end{abstract}

\maketitle

\section{Introduction}

\begin{figure*}[t]
\includegraphics[width=2\columnwidth]{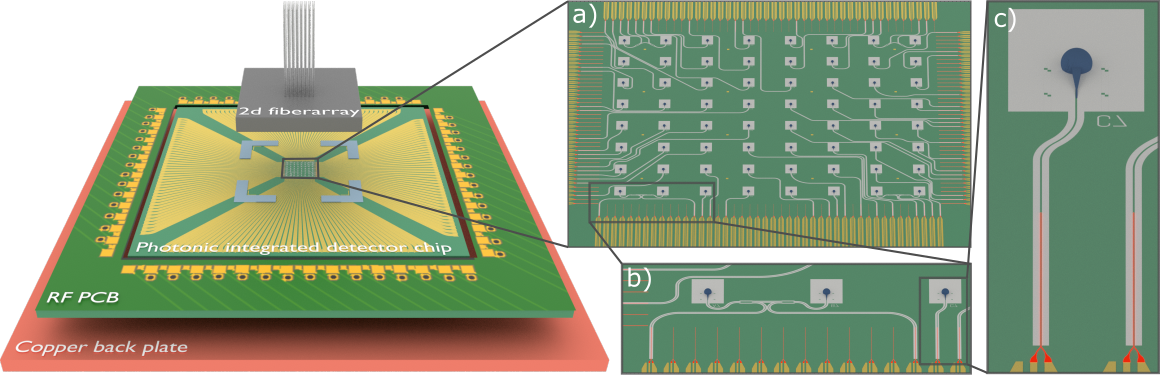}
\caption{\label{fig:chipschematic} Detector module and chip design. The superconducting detector module consists of a patterned silicon chip mounted on a copper back plate for thermalisation in a cryogenic environment. Electrical access to co-planar waveguides on the chip (yellow) is established through an interposer board via wirebonds. The chip is optically accessed from the top through a 64-channel optical fiber array. A ridge made from SU-8 photo resin (light blue) surrounds the photonic structures in the chip center. a) Zoom onto the center of the chip with 3D fiber-to-chip interfaces, photonic structures and superconducting detectors. b) In addition to regular detector channels, the chip design comprises structures for aligning the 2D fiber array with respect to the matrix of coupling interfaces via optical feedback. c) For a regular detector channel, light coupled onto the chip via a 3D coupling interface (blue) is guided along a photonic waveguide to the sensing region, i.e. the superconducting nanowire (red).}
\end{figure*}

Scalable and efficient manipulation and detection of single-photons are essential prerequisites for advancing the development of quantum technologies such as quantum key distribution or optical quantum information processing\cite{diamanti2016practical, slussarenko2019photonic, o2009photonic}. Integrated photonics has emerged as one of the most promising approaches for implementing large numbers of corresponding optical components with excellent interferometric stability as desired for processing quantum information encoded in single photons \cite{moody2022roadmap}. Nanofabrication methods originally developed for semiconductor thin film processing have allowed to realize essential functionalities for manipulating optical qubits in waveguide-networks on monolithic silicon chips\cite{qiang2018large}. Respective active and passive nanophotonic elements on silicon chips have shown similar performance to their bulk optic equivalents, thus enabling complex but highly integrated quantum optics experiments\cite{silverstone2016silicon,wang2020integrated} with the added benefits of outstanding interferometric stability and spatial mode control\cite{ma2014zeno}. 

Superconducting nanowire single photon detectors (SNSPDs) have developed into one of the most attractive photon counting solutions in quantum technology and offer exciting perspectives for a wide range of applications that demand ultimate photo-sensitivity\cite{zadeh2021perspective}. Importantly for integrated quantum photonics, high detection efficiencies over a wide spectral range, low noise, accurate timing, and high speed can also be achieved in waveguide-integrated SNSPDs\cite{ferrari2018waveguide}. 

An important challenge for future implementations of quantum technology is to scale current SNSPD solutions to larger numbers of individually addressable elements with customizable performance. Great progress in this regard has been achieved with superconducting detector arrays, where most state-of-the-art devices feature 16 to 64 elements\cite{doerner2017array_16,sinclair2019array_16,allmaras2020thermal_16,yabuno2020array_16,miki201464,allman2015array_64,miyajima2018array_64} but even a kilopixel array\cite{wollman2019kilopixel} has been demonstrated. In all of these implementations all N elements in the respective SNSPD array receive input from one common optical channel ($1 \times N$ geometry). Such realizations, where detector elements are individually addressable electrically but not optically, offer attractive prospects for imaging applications. Many applications in optical quantum communication, such as massively parallelized quantum key distribution systems, however, require multi-pixel SNSPD arrays where each detector element interfaces with a different, independent optical channel ($N \times N$ geometry). This can, for example, be achieved by positioning an optical fiber array above carefully aligned, meander-shaped SNSPDs \cite{zhang2021sixteen}. If additional processing, state analysis, or other manipulation of optically encoded quantum information is desired, it is further beneficial to exploit photonic integrated circuit technology for re-configurable application-specific functionalities and combine them with waveguide-integrated SNSPDs in a compact package\cite{beutel2021detector,gyger2021reconfigurable}. Such multi-channel chip-scale SNSPD solutions pose challenges in realizing efficient and mechanically stable optical interfaces with temperature independent performance as well as parallel electrical signal processing and readout at cryogenic temperatures\cite{marchetti2019coupling,shibata2019waveguide,wolff2021broadband}. The higher integration density as compared to stand-alone SNSPDs further makes it highly challenging to reach performance benchmarks set with devices that were optimized for individual operation, not the least because of increased thermal load, more complex fabrication processes and demanding packing solutions.   

Here we show how 64 waveguide-integrated SNSPDs on a silicon chip can efficiently be interfaced with an $8 \times 8$ channel 2D-optical fiber array for parallelized operation in the telecommunication C-band. Each detector element is individually addressable both optically and electrically and approaches the performance of stand-alone devices in primary benchmark disciplines. Our fiber-to-chip interfaces do not require sophisticated nanopositioning solutions at cryogenic temperatures but use low temperature compatible epoxy to hold the 2D-fiber array within the alignment tolerance of 3D-coupling structures. The latter are produced in direct laser writing of a photo-resin and provide efficient out-of-plane coupling between optical fibers and nanophotonic waveguides that guide light to preselected superconducting nanowire single-photon detectors\cite{gourgues2019controlled}. A custom-designed hybrid cryogenic-room-temperature signal amplification chain ensures low temporal jitter for the electrical SNSPD-readout and facilitates high count rates by avoiding latching behavior. We show state-of-the-art SNSPD performance for 38 out of 64 fiber-coupled detectors when operating a rack-mounted closed-cycle He4 cryostat at 3.6\,K temperature. Our system enables practical and reliable optical access to nanophotonic circuits with waveguide-integrated superconducting single-photon detectors on up to 64 fiber optic channels.

\section{Design considerations}
\label{subsec:Design Considerations}

\begin{figure*}[t]
\includegraphics[width=2\columnwidth]{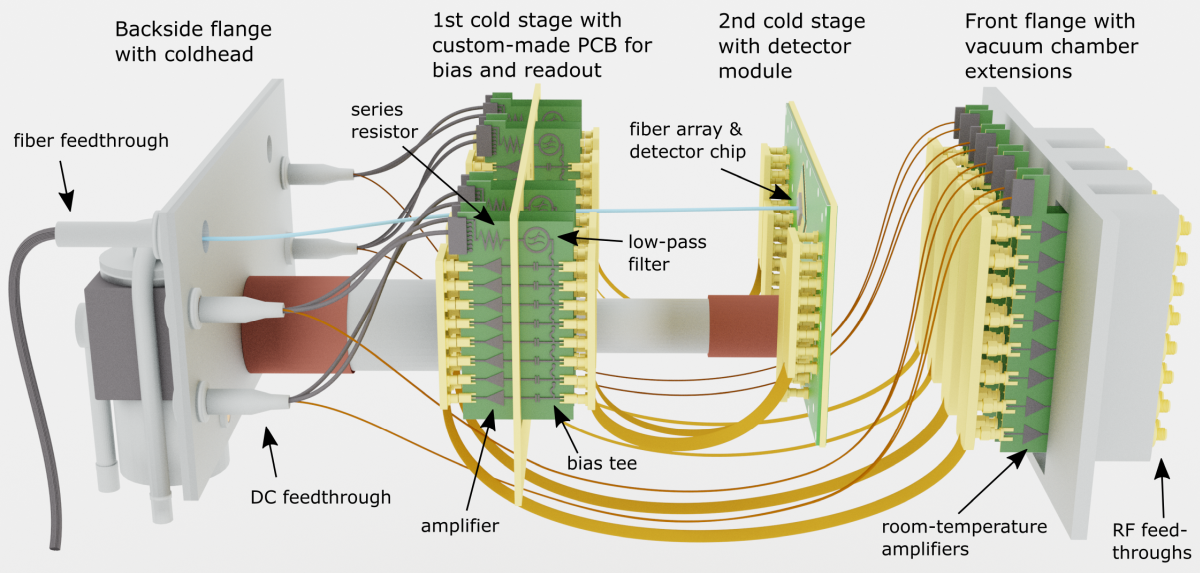}
\caption{\label{fig:compact_cryo_setup} Schematic of the inside of the 19"-rack-mountable cryostat. From left to right: The cryocooler is mounted on the backside flange of the cryostat, which also features hermetic electric multi-pin DC connectors and a custom-made 64-channel fiber optic feedthrough. Custom-made multi-layer PCBs mounted on the 1st cold stage host one cryogenic signal pre-amplifier and detector biasing electronics comprising series resistor, low pass filter and bias tee per channel. The packaged detector module is attached to the the 2nd cold stage. At the front flange, a custom-made room-temperature amplification stage is installed inside extensions made to the inner vacuum chamber of the cryostat. Amplified detector signals are guided outside the cryostat via 64 hermetic SMA connectors for further processing. In the schematic, bare fibers inside the cryostat are shown in light blue. Electric DC cable connections from the backside flange to the biasing and readout electronics are shown in black. Cryogenic ribbon cables for RF signal transmission between the detector module and the electronics on the 1st stage, as well as between the 1st stage and the 2nd stage are shown in orange. Note that at the 1st and 2nd cold stage RF connections are made using smooth-bore SMP connectors whereas at the room-temperature flange regular SMA connectors are employed.}
\end{figure*}

The principle design objectives consist in realizing a single-photon receiver system with a large number of detectors that can individually be addressed both optically and electrically. The receiver system should provide state of the art single-photon counting performance for a very wide range of use cases in sensing applications, such as Lidar, optical time domain reflectometry (OTDR), or lifetime imaging, and quantum technology, such as quantum key distribution, boson sampling, or quantum imaging. For many practical use cases it is highly desirable to provide a compact and portable receiver unit, which poses a challenge for SNSPDs that require corresponding cryogenic solutions. The receiver system should further interface with a time correlated single photon counting (TCSPC) unit that allows for parallel data acquisition on all detector channels.

We address these requirements by designing NbTiN SNSPDs in hairpin-shape, which can be fabricated in large numbers on a single silicon chip. While here we focus on a 64-channel SNSPD array, with detectors pre-selected from a total of 176 fabricated nanowires to guarantee high device yield, the approach readily scales to larger numbers of detectors. The chip is brought into thermal contact with the cold head of a closed-cycle He4 cryostat, and is mounted on a printed circuit board to facilitate electrical access to each individual detector, as shown in Figure \ref{fig:chipschematic}. The custom-designed cryostat fits into standard 19-inch racks and relies on a Gifford-McMahon cryocooler with a compact and cost-effective air-cooled compressor unit (see SI in the supplementary information for further details). Due to its compact size the system is portable and has successfully been used in field tests, which are reported elsewhere \cite{Terhaar2022QupadQKD}. 

Optical access to the SNSPDs on chip is provided by a 64-channel optical fiber array, see Figure \ref{fig:chipschematic}, with single-mode fiber-cores organized on an $8 \times 8$ grid featuring sub-micrometer positioning accuracy. As the positions of the 64 pre-selected SNSPDs (out of the 176 total) are only determined after nanowire fabrication, we integrate the selected hairpin nanowires with subsequently fabricated nanophotonic waveguides that allow for routing photons from 64 fixed optical interconnects, which align with the channels in the optical fiber array, to selected SNSPD locations, as shown in Figure \ref{fig:chipschematic}. The SNSPD length is here optimized for high absorption efficiency of photons traveling inside the waveguide, which results in somewhat slower reset dynamics due to increased kinetic inductance but benefits high detection efficiency. The interconnects are produced in 3D direct laser writing from a polymer resist and allow for efficiently coupling photons from an optical fiber with large mode field diameter to a nanophotonic waveguide with about an order of magnitude smaller mode field diameter. We here chose to demonstrate receiver performance for telecom wavelengths, i.e. 1550\,nm, which is a particularly interesting case for quantum communication and remote sensing applications but the optical interconnects achieve broad bandwidth through combining total internal reflection and quasi-adiabatic tapering. Due to the broadband absorption characteristics of superconducting nanowires, our results should thus readily extend to applications in the visible and near-infrared spectral ranges\cite{wolff2021broadband}. The fibers fed through the base of the cryostat can be connected to optical intensity monitors for calibration purposes as well as light sources outside the cryostat.

Electrical access to all 176 fabricated SNSPDs is possible via microstrip lines for pre-characterization purposes. In the packaged device only the 64 selected SNSPDs will be wirebonded to electrodes on the printed circuit board, from where flexible coaxial ribbon cables connect to bias tees mounted on the 50K-stage, as shown in Figure \ref{fig:compact_cryo_setup}. The dc-port of the bias tee allows for supplying current to the SNSPDs and connects to a stable multi-channel source outside the cryostat (see SII of the supplementary information), which also supplies up to 5\,V for all amplifiers inside the cryostat. 
The design of the electrical signal amplification and read out circuits that connect to the ac-port of the bias tee, on the other hand, has to satisfy several competing considerations. We target optimal timing accuracy and sustaining operation at high count rates for each individual channel while keeping the receiver in a compact, economic and portable format. As detector jitter is typically dominated by electronic noise in implementations relying on signal amplification at room temperature, we develop readout circuitry that combines cryogenic pre-amplification with subsequent signal amplification at room temperature for each channel. Observing requirements for compactness of the overall system, we integrate all 64 cryogenic pre-amplifiers on the 50K-stage and also mount 64 additional power-amplifiers on a room temperature flange on the inside of the cryostat, as shown in Figure \ref{fig:compact_cryo_setup}. Minimal power dissipation of the cryogenic pre-amplifiers was an important design objective here, given the large number of amplifiers inside the cryostat, all of which add to the overall thermal budget. We here consider two different custom cryogenic amplifier designs: (i) a single-stage version featuring low power consumption of 5.4\,mW, and (ii) a high-gain dual-stage version, dissipating 10.8\,mW, derived from the same design that was used for the room-temperature power-amplifiers. Amplified detector signals are recorded either with an oscilloscope or a TCSPC unit.

\begin{figure}
\includegraphics[width=\columnwidth]{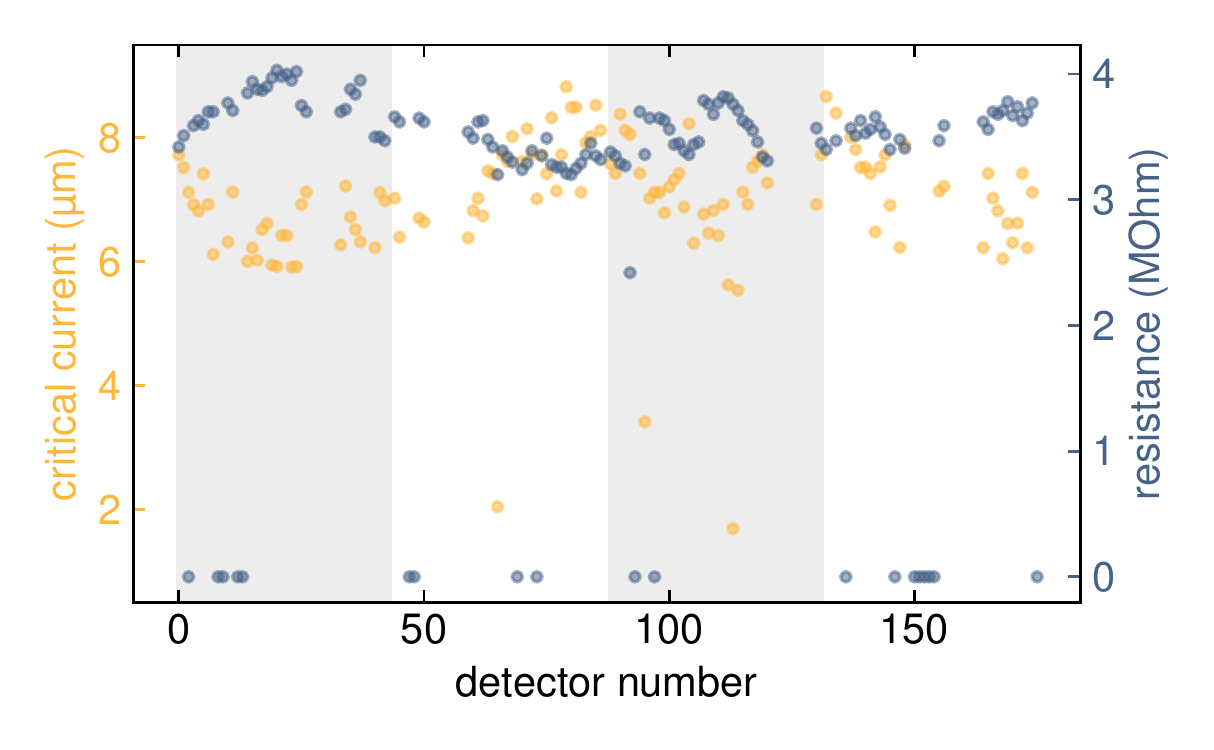}
\caption{\label{fig:maxSwC_R_vs_det} Critical current values at a temperature of 3\,K and room temperature resistance for a set of 176 fabricated NbTiN nanowires. Critical currents are measured in a single cooldown using a custom cryogenic measurement station. Missing data points indicate infinite resistance at room temperature for the respective nanowire.}
\end{figure}

\section{Chip fabrication and pre-characterization}
\label{subsec:Chip Fabrication and pre-characterization}

We fabricate waveguide-integrated SNSPDs from \SI{4.4}{\nano \meter} thin magnetron sputter-deposited NbTiN films on a \SI{340}{\nano \meter} stoichiometric silicon nitride (Si$_3$N$_4$) on insulator platform by combining electron-beam lithography (EBL) and optical lithography\cite{wolff2020superconducting}. The first layer consists of alignment markers and electrodes featuring impedance matching tapers\cite{zhu2019superconducting} and landing pads for electrical wire bonds. All structures are patterned in EBL, employing chemical semi-amplified positive-tone e-beam resist AR-P 6200 (CSAR) and physical vapor deposition of gold, followed by lifting off excess material in solvents. In the second layer, EBL is used to pattern U-shaped nanowires from a negative-tone hydrogen silsesquioxane (HSQ) resist mask that is precisely aligned to the marker structures in the first layer. Pattern-transfer from the mask into the NbTiN thin film is realized with reactive ion etching in CF$_4$ chemistry. A third layer utilizes EBL to cover the NbTiN nanowires with HSQ resist for protection from oxidation. At this point, our chip features 176 electrically accessible \SI{360}{\micro\meter} long and \SI{50}{\nano \meter} wide nanowires with a spacing of \SI{65}{\micro \meter}, arranged in a square pattern with 44 devices along each of the four edges of the square, as shown in Fig. \ref{fig:chipschematic}. In order to assess the suitability of each of these nanowires as detector elements for the final 64-channel waveguide-integrated SNSPD device, we determine both the room-temperature resistance using an electrical probe-station, and the critical current of each wire using a custom cryogenic measurement station equipped with  \SI{150}{\micro\meter} diameter Manganin wires for direct-current (DC) measurements on 150 individual lines. For the latter, we establish electrical connection from the chip to an interposer board via Al-wirebonds. The results of the measurements are depicted in Fig. \ref{fig:maxSwC_R_vs_det}. 

We observe position dependent variations of the room temperature resistance and critical current values, which we attribute to spatial inhomogeneity of the poly-crystalline NbTiN thin film\cite{thoen2016superconducting}. After accounting for layout-restrictions in photonic-waveguide routing, we select 64 representative nanowires with critical currents close to or higher than the mean critical current and room-temperature resistance close to the mean room-temperature resistance. In the photonic waveguide layer of the chip layout we route the waveguide sections associated with the 64 pre-selected nanowires via single-mode strip waveguides to 64 locations on a grid corresponding to the spacing between optical fibers in an $8 \times 8$ array, which will be positioned above the chip (see Fig. \ref{fig:chipschematic}). The nanophotonic layer further contains multi-mode interference splitters for calibration purposes and alignment markers for direct laser writing (see below). All structures are fabricated from the \SI{340}{\nano \meter} thin silicon nitride layer in EBL using CSAR-resist and reactive ion etching in CHF$_3$ chemistry. We found that it is essential to cover all large electrode structures during the etching process with (positive tone) resist to prevent contamination of the chip surface with Au residue. In an additional fifth layer, we use photolithography for patterning a \SI{30}{\micro \meter} high ridge from negative-tone epoxy-based photoresist (SU-8) that surrounds all photonic structures in the chip center. This ridge acts as a barrier for protecting the 3D fiber-to-chip interfaces, to be produced subsequently in direct laser writing, and preventing intrusion of light-absorbing cryogenic adhesive during the packaging process, described below. Lastly, the 3D fiber-to-chip interfaces are produced by two-photon polymerization employing direct laser writing in a photo-resin (IP-dip, Nanoscribe GmbH). Positioning of the \SI{26}{\micro \meter} high and \SI{110}{\micro \meter} long polymer coupling structures with respect to the Si$_3$N$_4$ waveguides is achieved using dedicated alignment markers in the nanophotonic layer. 

\section{Electrical and optical packaging}
\label{subsec:Packaging and optical transmission measurement}

The fabricated detector chip is mounted on a custom-designed printed circuit board (PCB) featuring a 30x30 mm opening for efficiently cooling the chip via a copper backplate, which is thermally ankered on the 3K-stage of the cryostat. The multi-layered PCB features 64 surface microstrip lines for routing electrical detector signals from signal and ground bonding pads to high-bandwidth SMP connectors. Electrical connections between on-chip electrodes and the PCB are made via Al wirebonds. On the other end, the SMP connectors interface with broadband and flexible cryogenic coaxial ribbon cables that route electrical signals from the PCB to a first stage of preamplifiers mounted on the 50K-stage of the cryostat. Subsequently, similar cables are used to direct the amplified signals to a second preamplification-stage, which too is integrated into the cryostat assembly but in direct thermal contact with outer cryostat hull at room temperature. The signals can be read out with an oscilloscope or a 64-channel TCSPC-device (MultiHarp 160, PicoQuant GmbH\footnote{https://www.picoquant.com/products/category/tcspc-and-time-tagging-modules/multiharp-160}). 

\begin{figure}
\includegraphics[width=0.9\columnwidth]{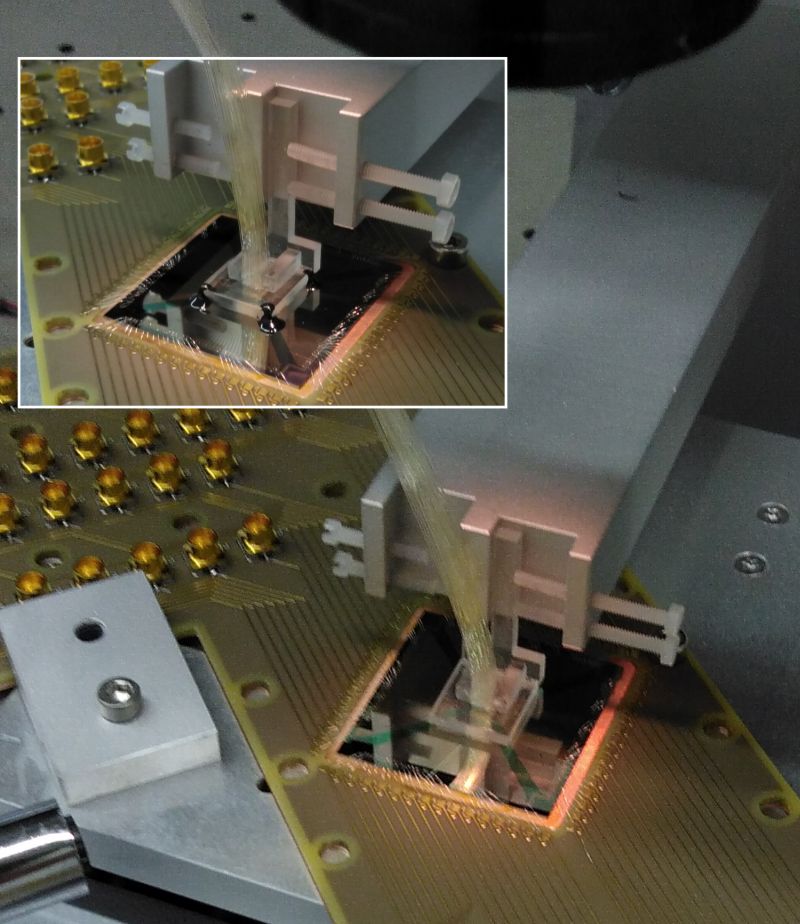}
\caption{\label{fig:glueing_packaging_photo_combined} Image of the packaging setup used for aligning the 2D fiber array to the on-chip coupling interfaces. The PCB with the chip is mounted on a six-axis translation stage while the fiber array is held at a fixed position. The process is monitored with a long working distance microscope (top, black) and a flexible USB microscope (side, silver). The inset shows the package after the epoxy is applied.}
\end{figure}

\begin{figure}
\includegraphics[width=0.9\columnwidth]{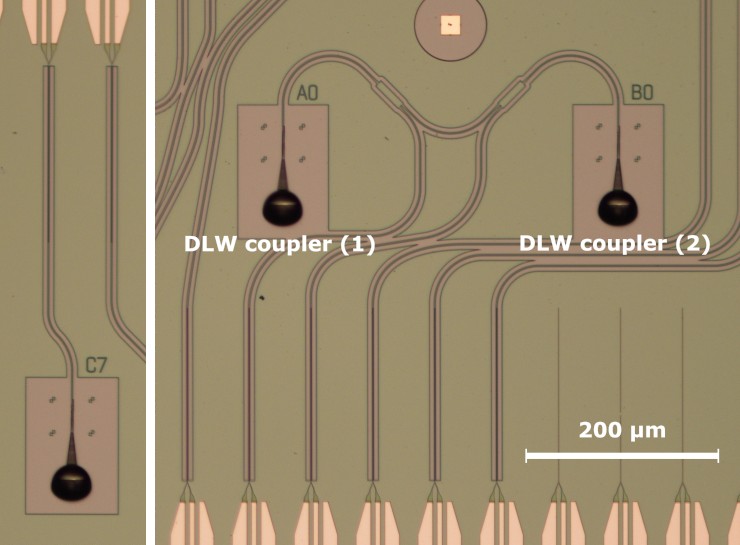}
\caption{\label{fig:optical_combined} Left: Optical micrograph of a regular detector device on the chip, with a nanophotonic waveguide connecting a DLW coupler (bottom) with a waveguide integrated SNSPD (top). Right: Dedicated structure for fine-alignment of the chip-PCB assembly relative to the fiber array, with a direct connection between two DLW couplers via two multimode interference devices (top).}
\end{figure}

Prior to mounting the chip-PCB assembly inside the cryostat we use a six-axis translation stage that allows for aligning the 2D fiber array to the on-chip coupling interfaces, both designed for a pitch of \SI{350}{\micro \meter} in x and y directions. The corresponding packaging setup is depicted in Fig. \ref{fig:glueing_packaging_photo_combined}. 
Coarse alignment of position, tilt and rotation of the chip-PCB assembly relative to the fiber array (fixed) is achieved utilizing a long working distance microscope and a flexible USB microscope. Fine alignment is then done by optimizing the optical transmission from a 1550\,nm wavelength CW laser source through dedicated devices consisting of a DLW coupler (1) for coupling light onto the chip, two $50:50$ multi-mode interference (MMI) splitters and a DLW coupler (2) for coupling back into an optical fiber of the array, as shown in Fig. \ref{fig:optical_combined}. We achieve optimal alignment when maximizing the transmission through two such devices at far corners of the chip optically connected in series. 

\begin{figure}
\includegraphics[width=\columnwidth]{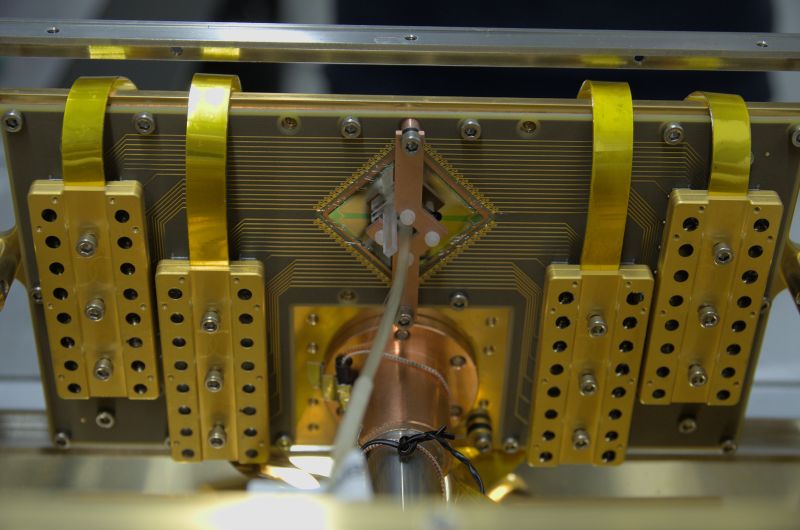}
\caption{\label{fig:3Kstage} Picture of the detector module on the 3K-stage of the cryostat with four attached 16-channel cryogenic coaxial
ribbon cables and fiber array mounts. Note that for operation the detector chip is additionally shielded from heat radiation via a housing from copper sheet.}
\end{figure}

The optimization procedure of the optical transmission requires adjusting the x,y and z positions together with the rotation of the chip relative to the fiber array. Note that a fiber polarization controller was employed before each input fiber to match the input polarization to the respective single-mode waveguide. Once maximum transmission is reached, the chip and fiber array are fixed in this position by applying a small amount of epoxy to each corner of the fiber array using a bare optical fiber\cite{mckenna2019cryogenic}. The inset of Figure \ref{fig:glueing_packaging_photo_combined} shows the package after the epoxy is applied. The epoxy, on the one hand, provides a mechanical connection between the array and the chip. On the other hand, it serves as a reference for the optimal coupling position. Once the epoxy has hardened, the assembly is mounted on the second stage of the closed-cycle cryostat, as shown in Fig. \ref{fig:compact_cryo_setup}. To ensure that the array is held in position during the cooldown process, it is additionally clamped into the reference position delineated by the epoxy markers  using a copper bracket equipped with polytetrafluoroethylene (PTFE) screws. A picture of the packaged detector module mounted at the cryostat's 3K-stage is given in Figure \ref{fig:3Kstage}. A custom-made 64-channel hermetic fiber feedthrough allows for optically accessing the chip from outside the cryostat.

\begin{figure}
\includegraphics[width=\columnwidth]{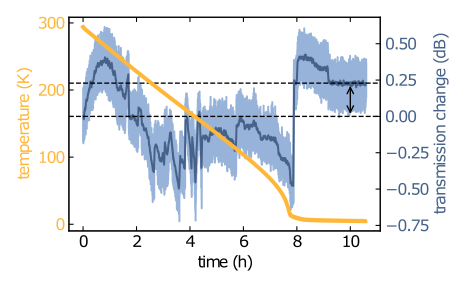}
\caption{\label{fig:time_temp_trans} Variations in optical transmission for the fiber-to-chip coupling during the cooldown of the cryostat. The temperature is measured at the second stage of the closed-cycle cryostat, starting at room temperature and reaching 3.6\,K after approximately 8.5\,hours. The change in optical transmission is shown relative to the transmission at room temperature, measured through four DLW couplers and two MMIs (see main text). At the base temperature of 3.6\,K, we find approximately +0.25\,dB higher transmission than at room temperature.}
\end{figure}

In order to quantify the stability of our setup against thermal drifts and deformations during the cooldown process, we monitor the transmission through the four DLW couplers and MMIs as done for the alignment procedure. Figure \ref{fig:time_temp_trans} shows how the temperature at the second cooling stage of the cryostat and the optical transmission behave over time during cooldown. We find a maximum variation in transmission of below $\pm\,$\SI{0.75}{\decibel} compared to the initial transmission at room temperature. At the base temperature of 3.6\,K we find stable transmission within approximately \SI{0.25}{\decibel} of the value at room temperature, thus confirming the excellent thermal stability of our assembly.

\section{SNSPD detection efficiency characterization}
\label{subsec:SNSPD detection efficiency characterization}

We pre-assess all 64 channels at the 3.6\,K base temperature of our compact closed-cycle cryostat for their optical and electrical connections as well as electrical transport through the superconducting nanowires. We find that 38 channels show proper electrical connection to the nanowire devices and response to optical input, thus forming functional detector-channels. A subsequent detailed analyses of the electrical and optical connections points to electrical connectivity problems of the SMA connectors at the second amplification stage inside the cryostat as a main source of failure. We conclude that the mechanical fabrication tolerances of the rigid blocks onto which the SMA connectors were mounted were too large to ensure precise matching between two rigid block assemblies, resulting in device failure. The SMP connections, on the other hand, perform reliably even at cryogenic temperatures and should be used in future hardware implementations. Other sources of failure include electrical solder connections, wirebonds, and in one case a broken optical coupler. For four non-functioning devices, we could not exclude damage to the nanowires during processing and packaging as a source of failure. While the overall channel yield was moderate in this implementation, less than 10\,\% of all devices showed defects that may be related to on-chip components. We anticipate that more than 90\,\% device yield is possible with straightforward improvements of electrical connections between the chip and the data acquisition system outside the cryostat.  

\begin{figure}
\includegraphics[width=\columnwidth]{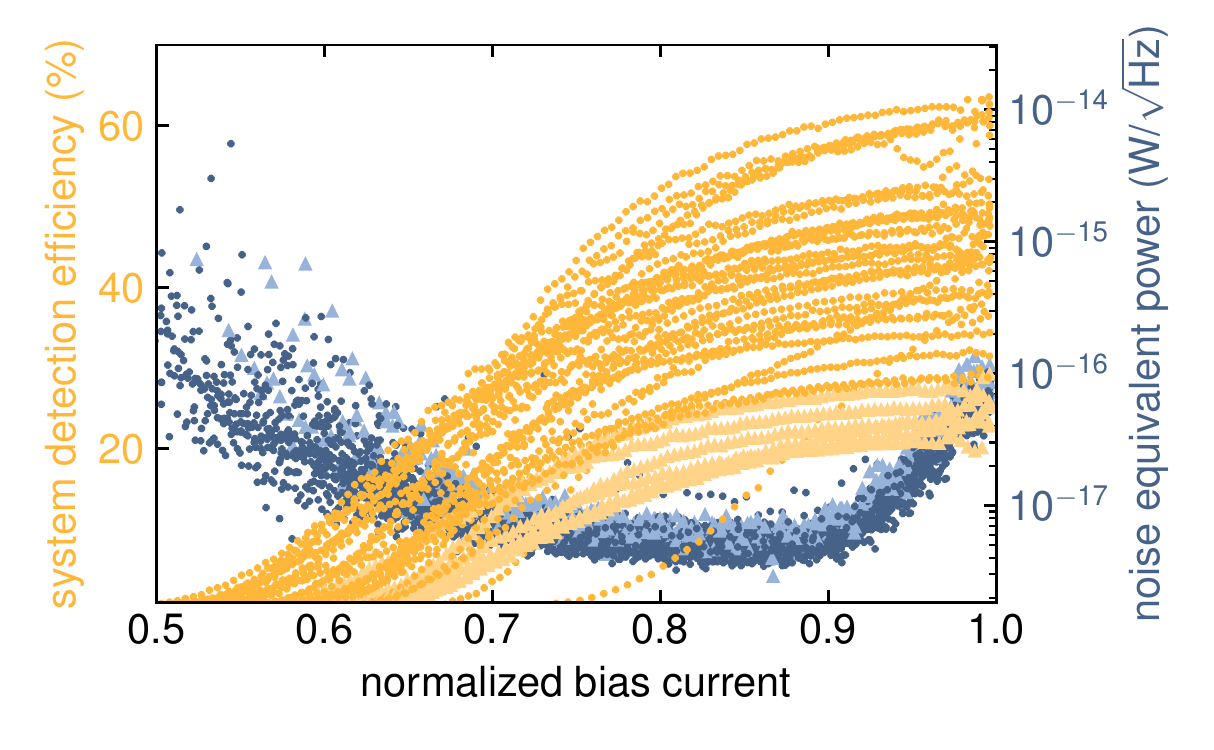}
\caption{\label{fig:detection_efficiency} System detection efficiency and noise-equivalent power of 38 waveguide-integrated SNSPDs in the multi-channel system measured at 3.6 K. For each detector, the bias current was normalized to the current setting where the dark count rate reaches 10\,\% of the total count rate at an input flux of $10^6$ photons per second. Lighter colors represent measured data for devices used for alignment purposes, i.e. containing an integrated MMI splitter that reduces the photon flux arriving at the detector by approximately 3\,dB.}
\end{figure}

For all operational channels we determine the individual system detection efficiency (SDE) using a cw laser source at a wavelength of 1550\,nm. We use calibrated variable optical attenuators, a fiber polarization controller and an optical power meter for setting the input photon flux to $10^6$ photons per second into the respective optical fiber of each channel. We employ a custom-designed 64-channel current source, which is described in further detail in SII of the supplementary information, for biasing each detector and measure the count rate as a function of bias current for each SNSPD. We find state-of-the-art system detection efficiencies for the waveguide-integrated SNSPDs of each channel, as shown in Figure \ref{fig:detection_efficiency}. Devices which feature $50:50$ MMI splitters for optical alignment purposes (see section \ref{subsec:Packaging and optical transmission measurement}) show system detection efficiencies of 20\,\% $<$ SDE $<$ 30\,\%, indicating that up to 60\,\% could be achieved when adjusting MMI splitting ratios accordingly or if specific channels would exclusively be dedicated for alignment purposes. Correspondingly, all other devices show 30\,\% $<$ SDE $<$ 60\,\%, which is significantly higher than most recent SNSPD array implementations \cite{allman2015array_64,miki201464,wollman2019kilopixel} and well suited for a wide range of demanding single-photon counting applications. The most significant contribution to both optical loss and variations in system detection efficiency will likely originate from the fiber-chip optical interconnects, as we selected nanowires of very similar properties to compensate for fabrication imperfections. Variable optical coupling efficiency may arise due to variations in the alignment of direct laser written structures to the nanophotonic waveguides, reproducibility variations of the 3D-polymer shapes and dielectric waveguide tapers, or other fabrication-related issues. A more detail discussion can be found in section SIII of the supplementary information. Additionally, an uncertainty analysis of the detection efficiency measurements is provided in section SIV of the supplementary information.

We further determine the dark count rate for each channel, when reducing ambient light to a minimum, thus avoiding exposure of SNSPDs to stray light, which would efficiently be coupled into nanophotonic waveguides due to the broad bandwidth of the DLW fiber-chip interconnects \cite{wolff2021broadband}. In Figure \ref{fig:detection_efficiency} we combine the corresponding dark count rates and system detection efficiencies into the noise-equivalent power (NEP) as one figure of merit for each channel. With NEP $< 5\cdot10^{-17}\,\mathrm{W}/\sqrt{\mathrm{Hz}}$, the noise-equivalent power of our system shows competitive performance across all 38 operational channel, benefiting, for example, remote sensing applications\cite{schuck2013otdr}. 

\begin{figure}
\includegraphics[width=1.0\columnwidth]{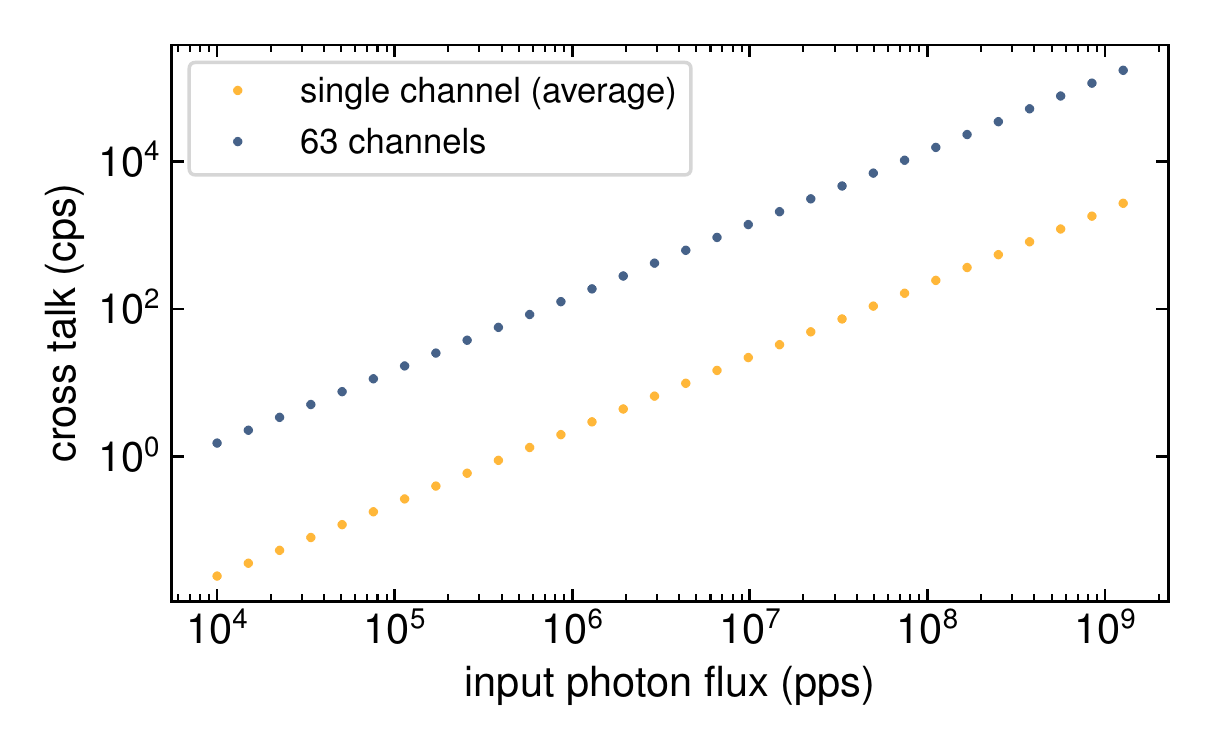}
\caption{\label{fig:cross_talk} Average crosstalk between two channels of the 64-channel detector module (yellow) and calculated cumulative crosstalk from 63 channels (blue) based on measurements for representative channels (see main text). We find a mean channel to channel crosstalk of around -60\,dB which results in an overall cumulative crosstalk well below -40\,dB.}
\end{figure}

Optical and electrical crosstalk between individual detectors is another important performance metric for highly integrated multi-channel single-photon receiver systems. We characterize such crosstalk for three representative devices (chip-center, -edge, in between), for which we record the count rate when biasing each detector at currents for which minimal noise equivalent power is reached, but with no optical input for the respective channel under test. We then provide an input photon flux of $10^6$ photons per second to all 37 other functional channels. On average we find crosstalk of -60\,dB per channel, corresponding to a cumulative crosstalk well below \SI{-40}{\decibel} for the channel under test when extrapolating to $64-1$ optically active channels, as shown in Figure \ref{fig:cross_talk}. Optical and electrical crosstalk thus plays a minor role for most applications, even when all channels receive optical input simultaneously. 

\section{SNSPD timing characteristics and electrical readout}
\label{subsec:SNSPD timing characteristics and electrical readout}

\begin{figure}
\includegraphics[width=1.0\columnwidth]{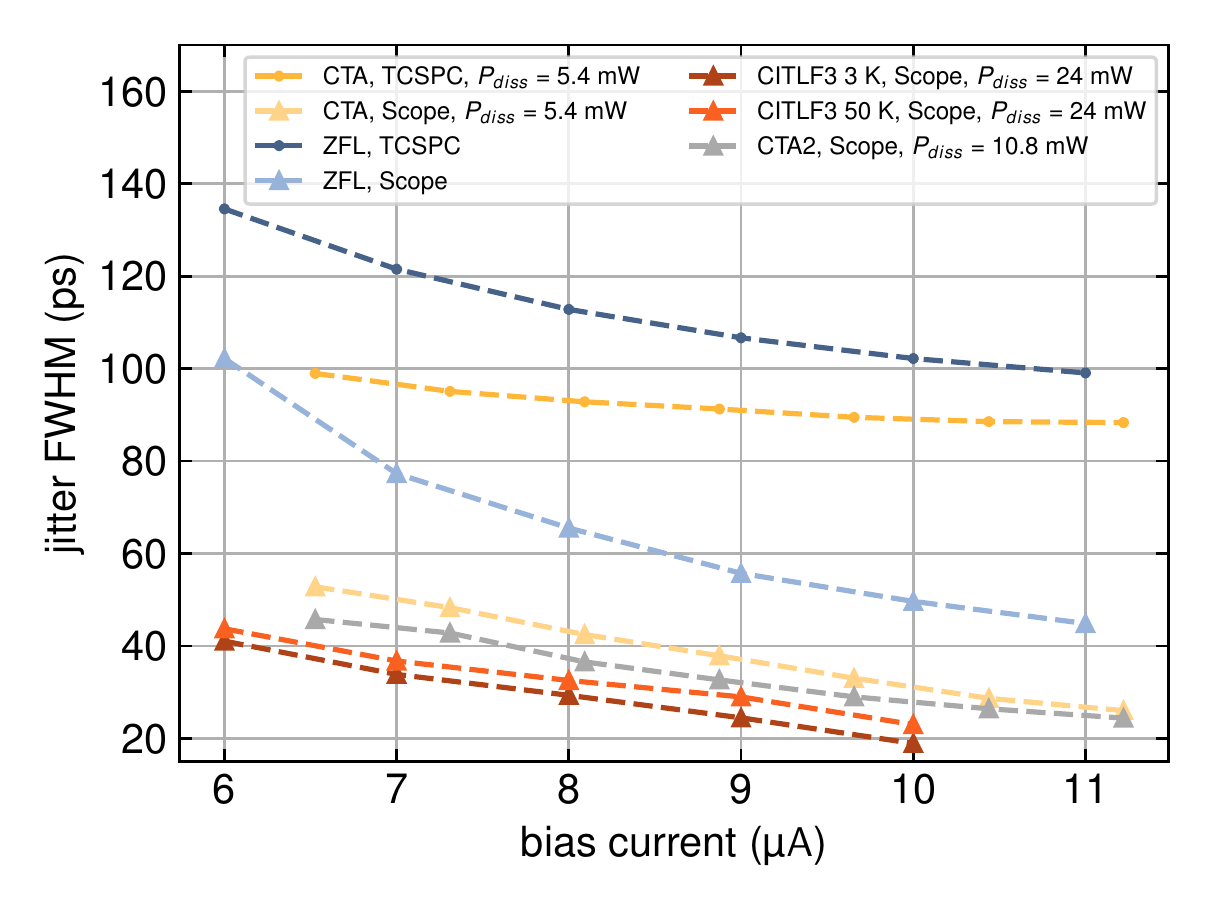}
\caption{\label{fig:jitter} 
The timing accuracy of four different preamplifier solutions is evaluated with an oscilloscope as well as a TCSPC system: one commercial solution operating at room temperature (ZFL), one commercial cryogenic solution (CITLF3) and two custom cryogenic preamplifiers, the first with a single-stage design (CTA) and the second with a dual-stage design (CTA2). With 5.4\,mW, the CTA operates at a dissipated power much lower than the commercial cryogenic solution, and thus provides improved scalability. The dissipation of the CTA2 is doubled, but the higher gain allows to achieve timing accuracy very close to the commercial cryogenic amplifier solution. For a detailed discussion of the timing performance see section E. Decreased maximum bias currents for the CITFL3 amplifiers likely stem from an impedance mismatch in the biasing network as compared to our well-matched custom cryogenic amplifier solutions.
}
\end{figure}

We characterize the timing performance of our receiver unit for a representative channel and infer from the relatively small critical current variations across all selected SNSPDs (see Fig. \ref{fig:maxSwC_R_vs_det}) that similar timing performance can be expected for other channels. We note that these characterization measurements were performed at lower operating temperature of 2.5\,K in a separate cryostat that allows easier access to individual channels, which is hardly feasible in the fully packaged receiver unit due to the high integration density. As compared to operation at 3.6\,K, which we discuss further below, lower temperatures allow for larger bias currents, thus benefiting SNSPD timing performance and signal-to-noise ratio, therewith constituting a best performance scenario. Following established procedures for determining SNSPD timing performance with a high bandwidth oscilloscope\cite{haussler2020amorphous}, we find minimal jitter values of 26.0\,ps at maximal bias currents of \SI{11.3}{\micro \ampere}, as shown in Figure \ref{fig:jitter} for the combination of our custom single-channel cryogenic pre-amplifier (CTA) mounted on the 50K-stage and a room temperature power-amplifier. When exchanging the single-stage version (CTA) with a dual-stage cryogenic pre-amplifier (CTA2), the jitter value improves to 24.4\,ps at maximal bias current (see Figure \ref{fig:jitter}), however at the expense of twice the power dissipation. 

To put these custom signal amplification solutions, which were optimized for low power dissipation, into perspective with state-of-the-art SNSPD readout techniques, we perform jitter measurements with commercial low noise amplifiers for both room temperature and cryogenic operation. Employing two cascaded low-noise amplifiers (ZFL-1000LN+, Mini-Circuits) with $>$ 20\,dB gain and 2.9\,dB noise figure, jitter values increase to 44.9\,ps at maximal bias current (see Figure \ref{fig:jitter}). For commercially available cryogenic amplifiers (CITLF3, Cosmic Microwave Technology, Inc.) with $>$ 35\,dB gain and $<$ 0.1\,dB noise figure, on the other hand, jitter values improve to 22.0\,ps when mounted on the 50K-stage, which improves to 18.9\,ps when mounted on the 3K-stage, as shown in Figure \ref{fig:jitter}. The latter performance however, is achieved at a power dissipation of 24\,mW, which poses significant challenges in managing the thermal load of a multi-channel receiver unit as it is more than four times higher than our custom solution (CTA). We further note that our cryogenic amplifier solution provides a well-matched integrated bias-network, more compact footprint and significantly reduced component cost as required for multi-channel solutions. Further details on the amplification schemes, including electrical signal shapes, jitter histogram envelopes, S21 curves and optimal operating current are given in sections SV-SVIII of the supplementary information.

While above measurements establish the potential of our detector-readout scheme for applications requiring accurate timing performance, simultaneous operation of large numbers of channels with high throughput is not practical with oscilloscopes for data acquisition. Instead, we here develop a multi-channel TCSPC unit, which features low dead time in order to sustain high photon counting rates, as desired in high data rate applications. Electrical tests of the TCSPC unit show an intrinsic timing uncertainty of <\,75\,ps FWHM per channel. A measurement of the total jitter in SNSPD output signals acquired with the multi-channel TCSPC unit yields FWHM-values of 88.3\,ps,\footnote{In this measurement scheme two channels are involved, hence the timing jitter baseline for the 64-channel TCSPC is <\,105\,ps FWHM} as shown in Figure \ref{fig:jitter}. This sets a baseline for the achievable timing precision of our single-photon receiver that is met for multi-channel operation, as evident from the curves flattening out at high bias currents. We again compare this performance, achieved with our custom cryogenic signal amplification solution (CTA), to established low noise amplifiers operated at room temperature, which achieve 99.1\,ps jitter when operated with the TCSPC-unit, as shown in Figure \ref{fig:jitter}. 

We have shown that the performance of our single-photon receiver unit can be optimized via the bias current conditions for the most relevant detector benchmarks, with detection efficiency and timing accuracy benefiting from high bias currents (see Figures \ref{fig:detection_efficiency} \& \ref{fig:jitter}), while lower bias currents result in reduced dark count rates and consequently lower $NEP$ (see Fig. \ref{fig:detection_efficiency}). This tunability of the operating conditions allows for application specific optimization of the receiver system. However, in order to provide a reference value for general purpose use, we assess the figure of merit\cite{hadfield2009review}, $H = \eta/(DCR \cdot \Delta t)$, for the fully packaged multi-channel system operating at 3.6\,K along with system jitter measurements recorded with the TCSPC unit for multiple channels using our custom single-stage cryogenic amplifiers (CTA). The corresponding multi-channel cryogenic amplifier PCB and its integration into the system, as shown in Figure \ref{fig:compact_cryo_setup}, is further described in section SIX of the supplementary information. The results in Figure \ref{fig:detection_efficiency_H} show that the jitter values slightly increased from the previous assessment at 2.5\,K due to the elevated temperature of the fully packaged device but remains around 110\,ps at 90\% of the bias current, where maximal $H-$values of $>10^7$ are achieved on all channels. Due to the strong bias current dependence of the dark count rate we expect it to be feasible to improve the figure of merit by about three orders of magnitude when using narrow band spectral filters\cite{shibata2015ultimate}.

\begin{figure}
\includegraphics[width=1.0\columnwidth]{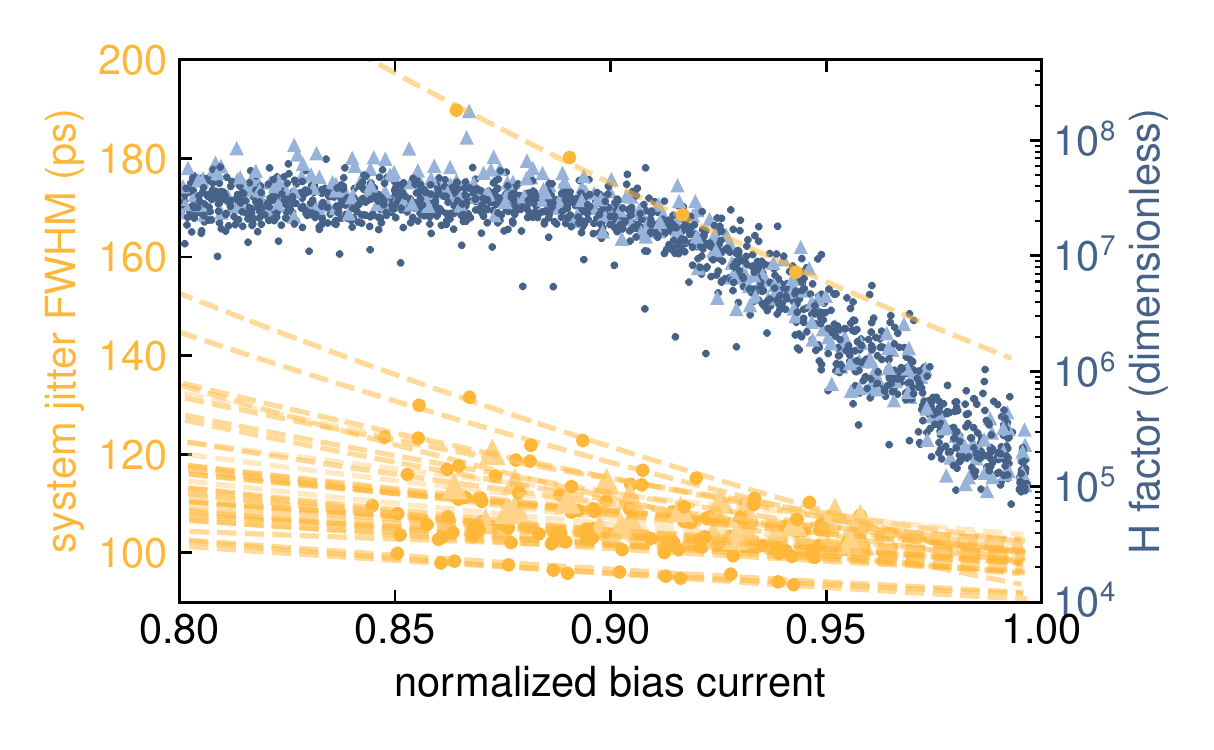}
\caption{\label{fig:detection_efficiency_H} Jitter and figure-of-merit, H, for 38 waveguide-integrated SNSPDs in the multi-channel system measured at 3.6 K. Lighter colors represent measured data for the alignment assist structures. Note that the graph contains data only for 37 detectors due to failure in the measurement procedure for a single detector.}
\end{figure}

\begin{figure}
\includegraphics[width=1.0\columnwidth]{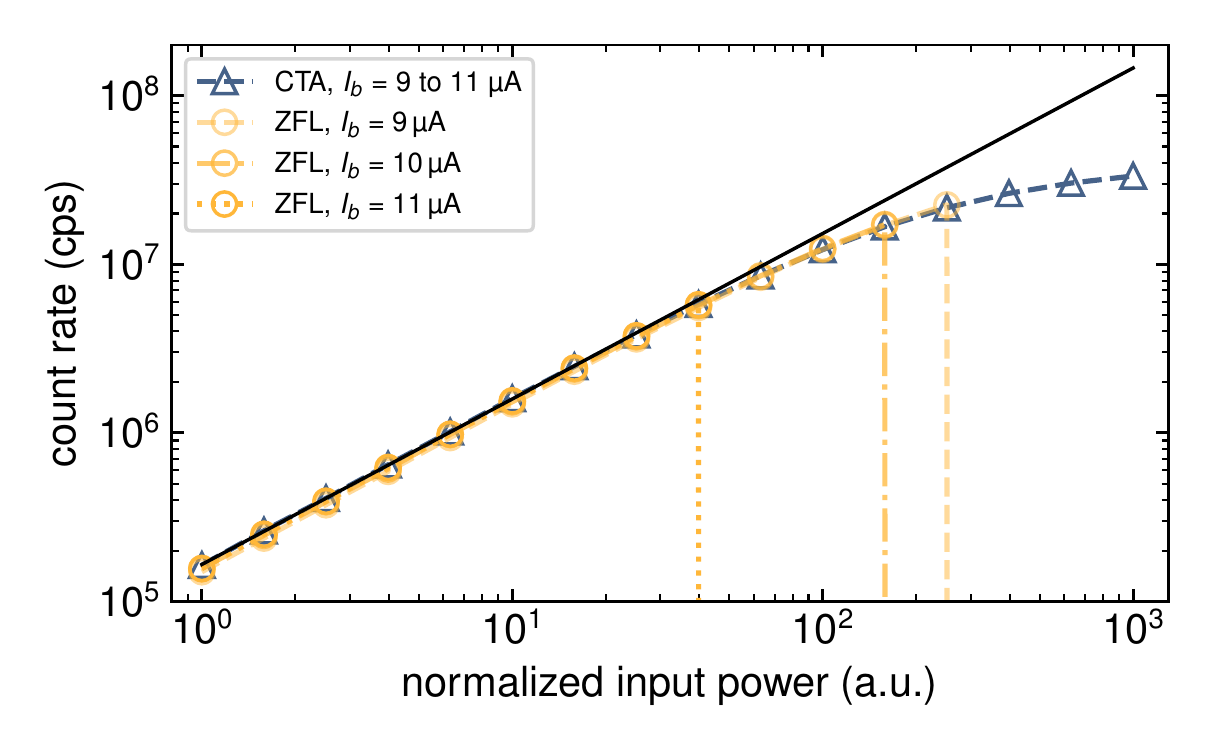}
\caption{\label{fig:countrate_versus_attenuation_10uA} Maximum detector count rate measured for different readout schemes at 2.5\,K and varying detector bias current. AC-coupling SNSPDs (ZFL, yellow) limits the achievable maximum count rate due to premature latching. This is caused by "over-biasing" the devices due to charging of the coupling capacitor that drives an additional bias current. Reducing the externally applied SNSPD-bias alleviates this problem and recovers maximum sustainable count rate to a certain extend but may have adverse effects on jitter and efficiency. The custom-made readout solution (CTA, blue) prevents such issues and count rates well above 20\,Mcps are achieved independent of the externally applied bias current. All measurements are recorded with the deadtime of the TCSPC system set to 20\,ns.}
\end{figure}

As a last performance parameter with practical relevance for applications, we evaluate the maximum count rate achievable with our receiver system. SNSPDs have shown promise for realizing single-photon counting with GHz rates\cite{vetter2016cavity} in specialized device geometries but in most general purpose implementations the maximum count rate is prematurely limited by the readout circuitry \cite{kerman2013readout, ferrari2019analysis}. Several techniques have been developed to tackle this electronic readout problem \cite{lv2018improving, zhao2014counting, lv2017large}, here we chose to follow an approach based on an L–R path to ground\cite{cahall2018scalable}, which allows to protect the SNSPDs from latching, i.e. a permanent transition to the normal conducting photon insensitive state. For a pre-characterization of the detector channels at 2.5\,K, we use a 1550\,nm wavelength cw laser source and illuminate the SNSPDs with systematically varied optical power. The resulting count rates are shown in Figure \ref{fig:countrate_versus_attenuation_10uA} for both our custom cryogenic amplifiers and commercial low noise amplifiers operating at room temperature. The latter exhibit bias-current dependent maximal count rates, while our readout circuit sustains larger count rates, indicating that the SNSPDs fully reset even for high input flux close and well above the inverse of the nanowire recovery time. Further data on the count rate at different bias conditions can be found in section SX of the supplementary information.

\begin{figure}
\includegraphics[width=1.0\columnwidth]{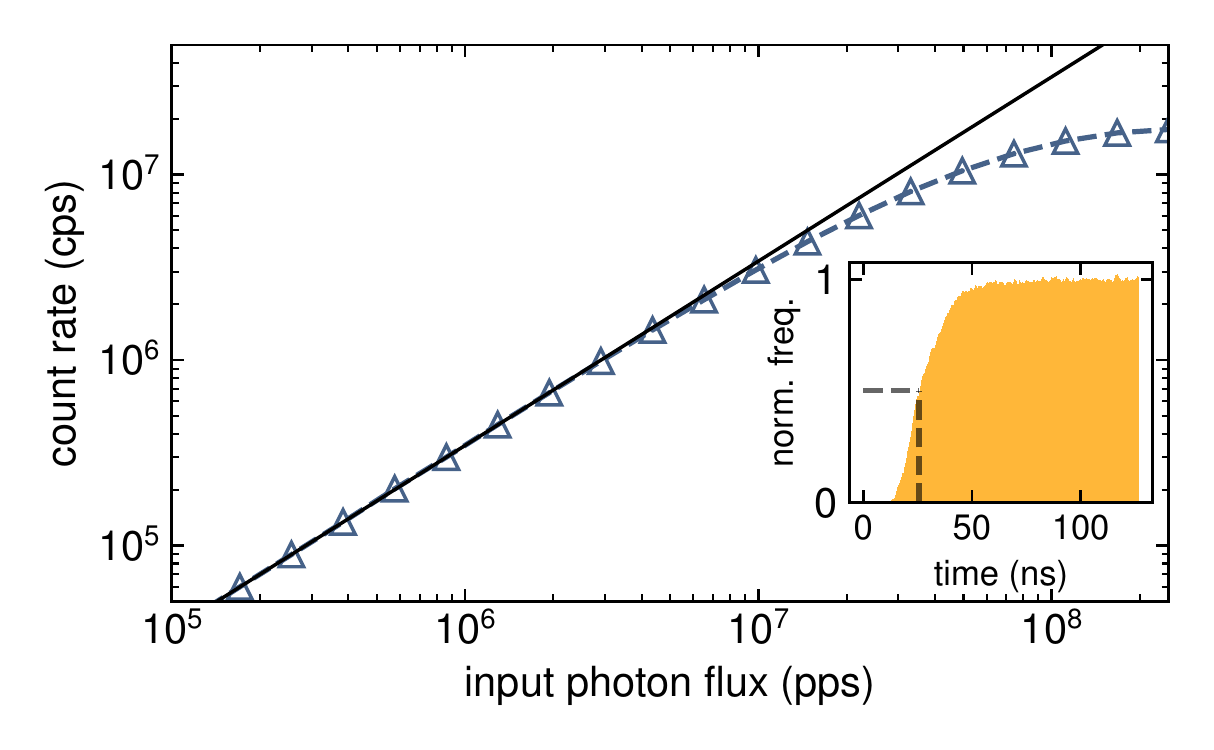}
\caption{\label{fig:iat} Count rate measurement of a representative detector device at 3.6\,K. Stable operation without detector latching is observed even at count rates of above 10\,Mcps. Inset: Measured count histogram for photon inter-arrival times. After a dead time of 10\,ns, the detection efficiency increases with time, recovers to half of the initial efficiency at 25\,ns and fully recovers at approximately 50\,ns.}
\end{figure}

In the fully packaged multi-channel receiver operating at 3.6\,K we find qualitatively similar behavior for individual fiber optic channels, as shown in Figure \ref{fig:iat}. We here further perform an inter-arrival time measurement for a single channel using a 1550\,nm wavelength cw laser source providing a constant input flux of $10^6$ photons per second. A histogram of the time intervals recorded between an initial detection event and subsequent detection events, i.e. a start-multi-stop measurement, is shown in the inset of Figure \ref{fig:iat}. The measurements reveal a dead time of 10\,ns in which no detection event can occur. After a time of 25\,ns the detection efficiency has recovered to 50\,\% of the maximum value and is fully recovered after 50\,ns. This performance is suitable to sustain count rates of at least 20\,MHz for a single detector. For larger input photon flux the count rate starts to deviate from expected behavior, as shown in Figure \ref{fig:iat}, which is a well-known effect of the dead time in single-photon counting experiments.

\section{Summary and Outlook}
\label{subsec:Summary and Outlook}

In this work we report on progress towards realising a multi-channel waveguide-integrated SNSPD array suitable for practical quantum technology and single-photon counting applications. Different from previous implementations, all detector elements in our receiver are individually accessible via fiber optic channels and electrical readout lines. Using a pre-selection technique, we are able to realize waveguide-integrated SNSPDs with high yield. High system detection efficiency of up to 60\,\% is achieved by interfacing SNSPDs with high internal quantum efficiency via low loss waveguides and polymer-based optical interconnects to a 64-channel optical fiber array. Variations in system detection efficiency between channels primarily originate in fabrication imperfections of the optical interconnects between optical fibers and nanophotonic waveguides. We here anticipate that further optimization of the 3D coupling interface can lead to coupling efficiencies that consistently exceed 50\,\%. We further developed an ultra-compact 64-channel electrical readout circuit, which is completely integrated into the cryostat, thus providing an attractive packaging solution that allows the system to be portable. High timing accuracy is achieved by employing custom designed cryogenic signal amplification circuits, which provide a good compromise between sub-30\,ps jitter performance and low power dissipation, as desired for multi-channel detector systems. The electrical signal processing chain present a clear point where improvements are needed as 25 out of 64 channels showed faulty electrical connections. We anticipate that these issues can be solved by designing more robust electrical interfaces, e.g. employing SMP instead of SMA connectors among other improvements. Parallel acquisition of up to 64 detector signals is possible with a TCSPC-unit, which sustains up to 20\,MHz count rates per channel at maximal detection efficiency and 110\,ps of overall system jitter.  

The final transportable prototype features 38 working channels in a closed-cycle cryostat operating at 3.6\,K, with a mean detection efficiency of 40\,\%, dark count rates of around 200\,Hz, and a channel-to-channel crosstalk of around -60\,dB. Improvements of the dark count rate of several orders of magnitude seem realistic when employing cold fiber filters or waveguide-integrated spectral filters. The freedom in designing the nanophotonic circuits, SNSPD geometries and electrical readout lines on a channel-by-channel basis will allow for realizing receiver units that feature individual channels with customized performance, such that some channels may be optimized for detection efficiency, while others are optimized for timing accuracy, etc. The excellent performance characteristics and prospects for channel-specific optimization of our multi-channel single-photon receiver system are ideally suited for a wealth of applications in future sensing and quantum technology applications, such as quantum key distribution\cite{Terhaar2022QupadQKD}. 

\begin{acknowledgments}
The authors acknowledge the financial support from the Federal Ministry of Education and Research of
Germany in the framework of project QuPAD (BMBF 13N14953). C.S. acknowledges support from the Ministry for Culture and Science of North Rhine-Westphalia (421-8.03.03.02–130428). H.G. thanks the Studienstiftung des deutschen Volkes for financial support. The authors thank Doreen Wernicke for support with the cryogenic system. We further thank the Münster Nanofabrication Facility (MNF) for their support in nanofabrication-related matters.  
\end{acknowledgments}

\bibliography{bibl}

\newpage
\textbf{\huge{Supplementary \\ information:}}
\setcounter{section}{0}

\section{Compact cryostat}

\begin{figure}
\includegraphics[width=\columnwidth]{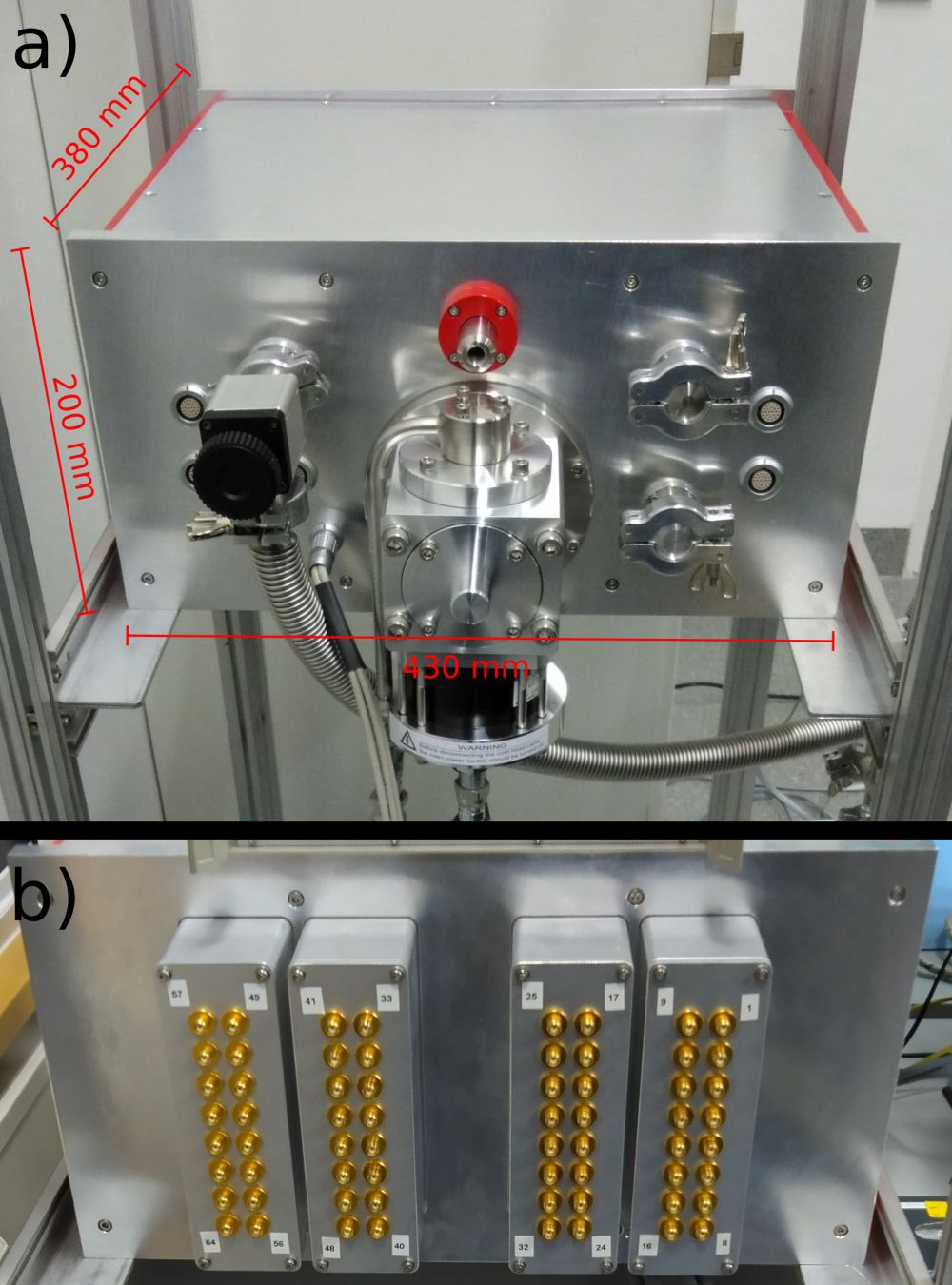}
\caption{\label{fig:cryostat} a) Back side of the 19"-rack-sized cryostat with cryocooler and flanges for pumping, fiber array feedthrough installation and electrical DC feedthroughs. b) Front panel of the cryostat with extensions hosting room-temperature amplification stages and 64 RF SMA feedthroughs.}
\end{figure}

A picture of the employed compact cryogenic system is shown in Figure \ref{fig:cryostat}. The cryostat's dimensions allow for practical integration into a standard 19" rack. The backside flange hosts the cryocooler (Sumitomo RDK-101DL) and vacuum pumping port, as well as KF16 ports used for a fiber feedthrough. Furthermore, hermetic DC feedthroughs allow for detector and amplifier biasing and temperature sensor readout. At the front flange, the room-temperature electric amplifiers are mounted in four rectangular extensions of the vacuum chamber. The amplified signal is read out via 64 individual hermetic SMA feedthroughs arranged in four 2x8 arrays. Access to the inside of the cryostat is provided through removable front and back side flanges as well as large side panels. The rectangular base plate at the cryocooler's first stage features four openings in which the 16-channel cryogenic amplifier modules are installed. A large encasement from gold plated copper sheet shields the second stage from heat radiation. The second stage comprises a large cold plate providing enough space for the detector module. 

\section{Powersupply for the preamplifiers and SNSPD bias supply}

\begin{figure}
\includegraphics[width=\columnwidth]{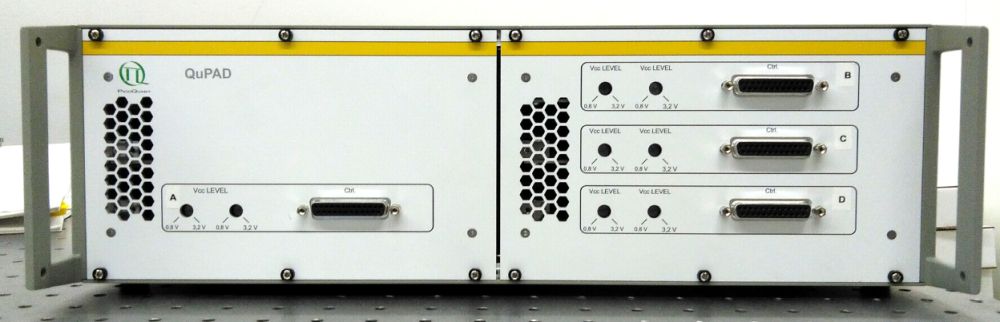}
\caption{\label{fig:bias_box} Multi-channel source for powering the amplifier chain and biasing the SNSPDs.}
\end{figure}

In order to bias the 64 room-temperature amplifiers, cryogenic amplifiers and SNSPDs, we develop a multi-channel bias supply. A picture of its front panel is given in Figure \ref{fig:bias_box}. Via potentiometers at the front panel, the voltage for both the room-temperature amplifiers and the cryogenic amplifiers is set manually to a value between 0 and 5\,V. Note that the amplifiers on an eight-channel PCB are powered via a single supply line. This offers a good compromise between scalability and supply line count, while still keeping cross talk over the supply rail minimal. The SNSPD bias current is set from a host-PC via a serial interface for each SNSPD individually. Electrical connection is established through four shielded 25 pin SUB-D connectors at the front panel. 

\section{SNSPD detection efficiency variation analysis}

\begin{figure}
\includegraphics[width=\columnwidth]{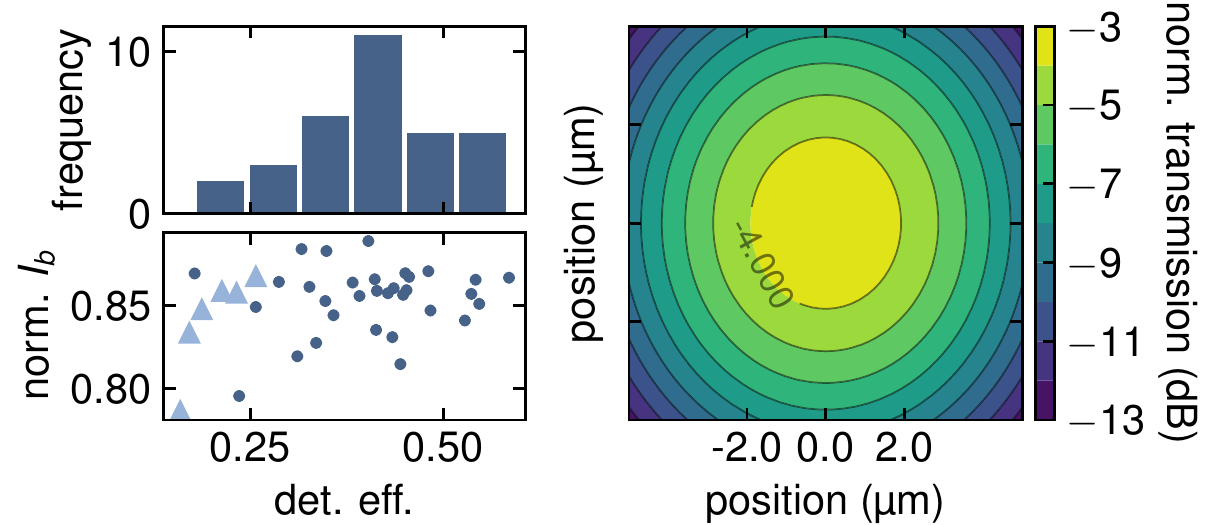}
\caption{\label{fig:de_nep} Left bottom: Detection efficiency at minimal noise-equivalent power. Lighter colors represent measured data for the alignment assist structures. Left top: Histogram of the measured detection efficiencies at minimal noise-equivalent power. Note that for the histogram the data of the alignment assist structures was excluded. Right: normalized transmission calculated for a mean polymer coupler device according to measurements of 88 polymer coupling devices.}
\end{figure}

In order to derive the variation of the detection efficiency, in Figure \ref{fig:de_nep} we plot the detection efficiency at the operating current, which we define as the current at minimal noise-equivalent power. We find a mean detection efficiency of the regular devices of 40.5\,\% with a standard deviation of 9.4\,\%. The relative deviation of $\Delta DE=0.23$ can be approximated by
\begin{CEquation}\label{equ:de_var}
\Delta DE_{calc} = \sqrt{\Delta C^2 + \Delta FAC^2 + \Delta FCO^2},
\end{CEquation}
where $\Delta C$, $\Delta FAC$ and $\Delta FCO$ denote the relative deviation in the transmission of the coupling structures, of the fiber array channels, and induced by spatial misalignment in the fiber-to-chip interface, respectively. 

We determine $\Delta C$ by measuring the transmission for 1550 nm wavelength light of devices made from two coupling structures connected by a waveguide on a reference chip at room-temperature. The measured sample of 44 of these devices, comprising in total 88 coupling structures, reveals a relative deviation in the transmission of $\Delta C = 0.17$. 

By measuring the transmission of each fiber array channel for 1550 nm wavelength light using a power meter we find $\Delta FAC = 0.024$. 

In order to determine $\Delta FCO$, we calculate the influence of the variation of the fiber core offset from the optimal position on the transmission variation. The variation in fiber core offset is given by the manufacturer and is \SI[separate-uncertainty = true]{0.85(48)}{\micro\meter}. In order to determine the influence of the offset, we measure the transmission as a function of the position of a fiber array relative to a coupling structure at optimal height. As a result we obtain a 2D plot, which is given in Figure \ref{fig:de_nep}. From a 2D Gaussian fit to the data, we extract a variation of the transmission due to the mean fiber core offset of $\Delta FCO = 0.094$. 

Inserting into Equation \ref{equ:de_var} yields $\sigma_{DE, calc}=0.20$. This value is 15\,\% smaller than the directly measured value for $\sigma_{DE}$. This discrepancies could be attributed to further misalignment induced by the cooldown procedure combined with slight differences in the detection performance of the detectors, i.e. the absorption efficiency and the internal detection efficiency. Also limited positioning accuracy of the 3D polymer couplers may contribute to the detection efficiency variation. We conclude that the variation in the detection efficiency for preselected detectors can substantially be decreased by further minimizing the largest contribution, the variation in the coupling efficiency of the fiber-to-chip interface, i.e. further improving the reproducibility of the coupling structures.

\section{Detection efficiency uncertainty}

The detection efficiency $\eta$ of the individual channels is calculated using the equation

\begin{CEquation}
    \eta = \frac{CR - DCR}{\phi}
\end{CEquation}

from the measured count rate $CR$, dark count rate $DCR$ and input flux $\phi$. With $DCR\ll CR$ and assuming that $CR$ and $\phi$ are dependent variables, the relative uncertainty in the detection efficiency is given by

\begin{CEquation}
    \frac{u(\eta)}{\eta} = \frac{u(CR)}{CR} + \frac{u(\phi)}{\phi},
\end{CEquation}

where u(x) is the absolute error of x. 

In the following the individual contributions are discussed.

\subsection{Count rate uncertainty}

In order to estimate the uncertainty in the count rate, we record a count rate versus bias current curve for each detector three times. We then calculate the mean value and the standard deviation of the count rate for every channel of the system at its optimal operating point, i.e. the bias current that gives the minimal noise equivalent power. Averaging over all channels yields the relative uncertainty in the count rate  

\begin{CEquation}
    \frac{u(CR)}{CR} = 0.7\,\%.
\end{CEquation}

\subsection{Flux uncertainty}

For setting the constant, low photon fluxes between 0 and $10^9$ photons per second, that are required for measuring the performance of the detectors, laser light is attenuated using a series of two variable optical attenuators. The required attenuation for setting the desired photon flux onto the detector is calculated from an initial measurement of the power emitted by a Santec TSL-710 cw laser source at a wavelength of 1550\,nm using a calibrated powermeter. With this method the relative uncertainty in the flux is given by

\begin{CEquation}
    \frac{u(\phi)}{\phi} = \sqrt{\left( \frac{u(LS)}{LS} \right)^2 + \left( \frac{u(PM)}{PM} \right)^2 + \left( \frac{u(AL)}{AL} \right)^2},
\end{CEquation}

with the relative uncertainty of the laser output power due to its stability over time $\frac{u(LS)}{LS}$, the relative uncertainty of the power measurement with the powermeter $\frac{u(PM)}{PM}$ and the relative uncertainty of the attenuation originating from a non-linear behaviour of the attenuators $\frac{u(AL)}{AL}$. 

\subsubsection{Laser stability}

\begin{figure}
\includegraphics[width=\columnwidth]{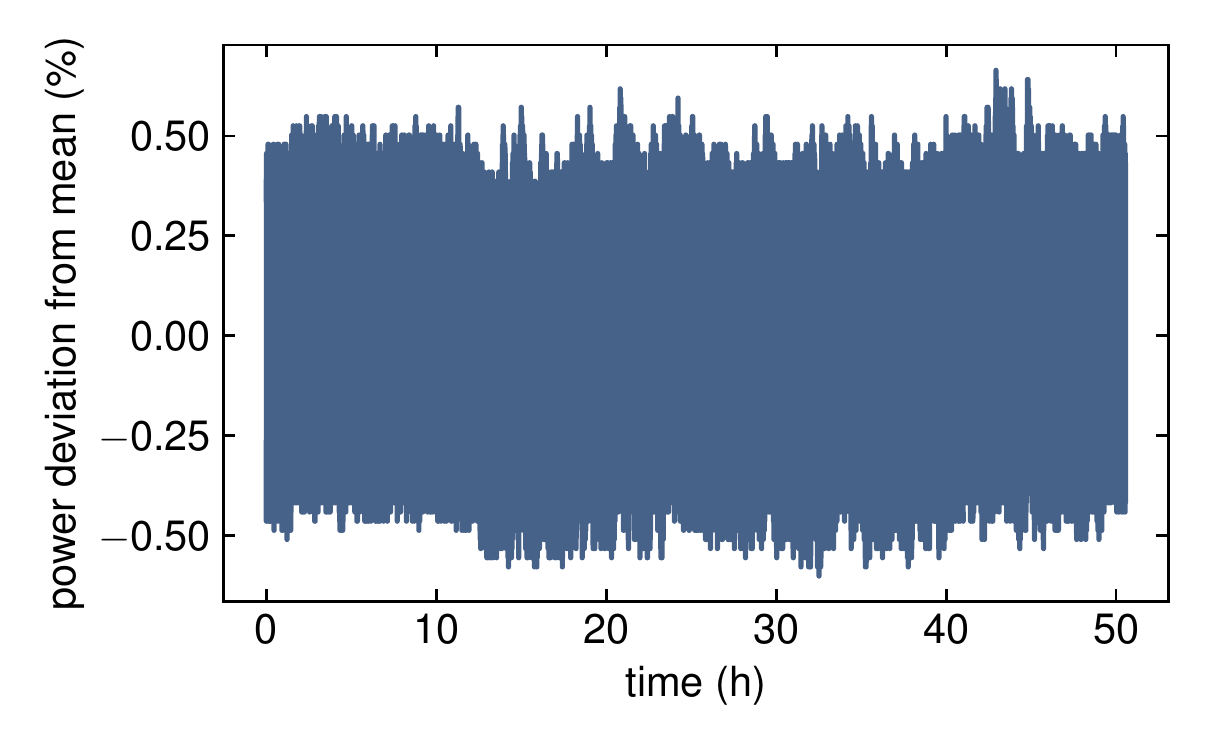}
\caption{\label{fig:laser_stability} Measurement of the laser stability over time. A standard deviation of below 0.3\.\% adds negligible error to the detection efficiency.}
\end{figure}

Careful measurement of the detector's efficiency in the multi-channel system takes several hours. We thus first characterize the laser source (Santec TSL-710) stability over time. The results of a long term measurement given in Figure \ref{fig:laser_stability} show an absolute deviation from the mean output power of $\frac{u(LS)}{LS}=0.3\,\%$ over a total time of more than two days.

\subsubsection{Powermeter uncertainty}

We extract a relative uncertainty of the Santec MPM-212 powermeter of $\frac{u(PM)}{PM} = 5\,\%$ from the device datasheet.

\subsubsection{Attenuator linearity}

\begin{figure}
\includegraphics[width=\columnwidth]{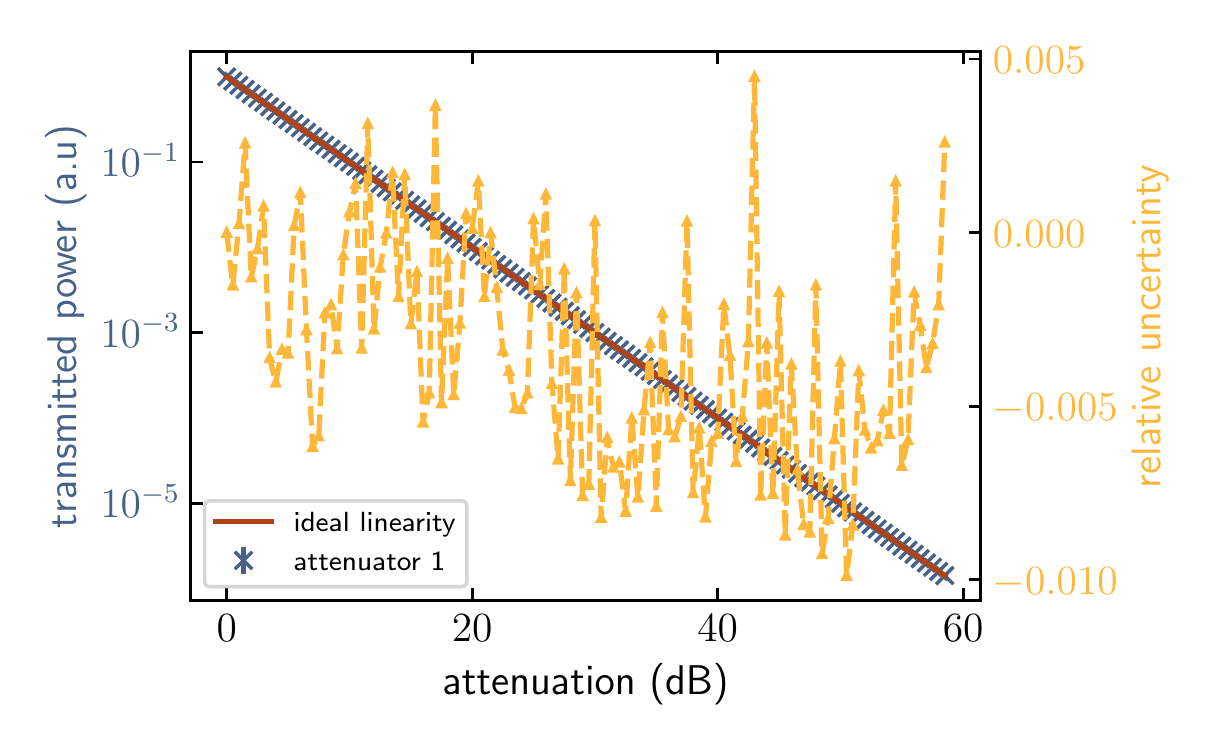}
\caption{\label{fig:attenuator_1_linearity} Normalized output power of the first variable attenuator (\textit{Agilent 81570A}) used for setting the photon flux onto the detector versus the set attenuation. The deviation from an ideal linear behaviour gives the relative uncertainty.}
\end{figure}

\begin{figure}
\includegraphics[width=\columnwidth]{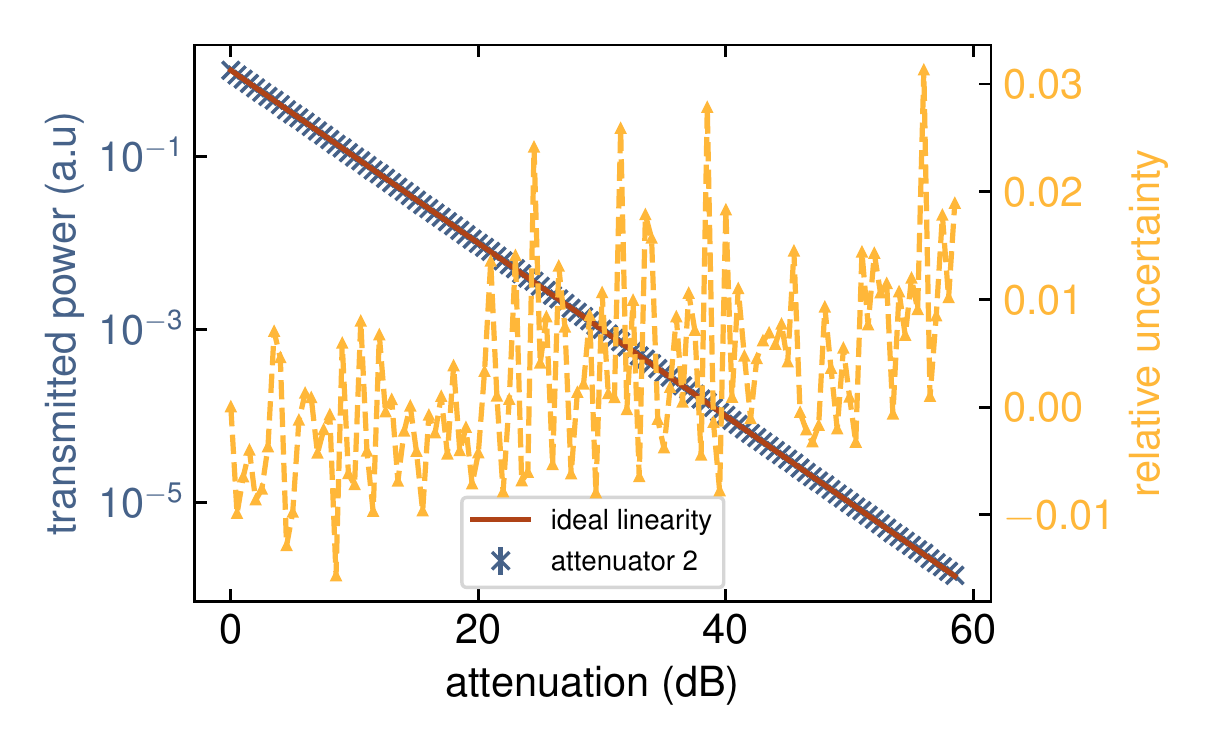}
\caption{\label{fig:attenuator_2_linearity} Normalized output power of the second variable attenuator (\textit{HP 8156A}) used for setting the photon flux onto the detector versus the set attenuation. The deviation from an ideal linear correlation gives the relative uncertainty.}
\end{figure}

In order to verify the linearity of the attenuators for different attenuations, we connect a cw laser source (Santec TSL-710) at a wavelength of 1550\,nm to the input of the first attenuator. The output of the first attenuator is connected to the input of the second attenuator and the output of the second attenuator is connected to a powermeter. We monitor the power while varying the attenuation of one of the attenuators from 0 to 60 dB while the other attenuator is set to 0 dB. The results of the measurement are given in Figure \ref{fig:attenuator_1_linearity} and \ref{fig:attenuator_2_linearity}. We find a maximum deviation from the optimal linear behaviour of 0.4\,\% and 3.1\,\% for the two different attenuators. The total relative uncertainty due to the non-linear behaviour of the attenuators thus is

\begin{CEquation}
     \frac{u(AL)}{AL} = \sqrt{0.004^2 + 0.031^2} = 3.1\,\%.
\end{CEquation}

\subsection{Total uncertainty}

With the values derived above for the individual contributions to the uncertainty in the detection efficiency, we find a total uncertainty in the detection efficiency averaged over all channels of

\begin{multline*}
    \frac{u(\eta)}{\eta} = \\ \frac{u(CR)}{CR} + \sqrt{\left( \frac{u(LS)}{LS}\right)^2 +\left( \frac{u(PM)}{PM} \right)^2 + \left( \frac{u(AL)}{AL} \right)^2} = \\ 6.6\,\%.
\end{multline*}

\section{Voltage traces for different readout schemes}

\begin{figure}
\includegraphics[width=\columnwidth]{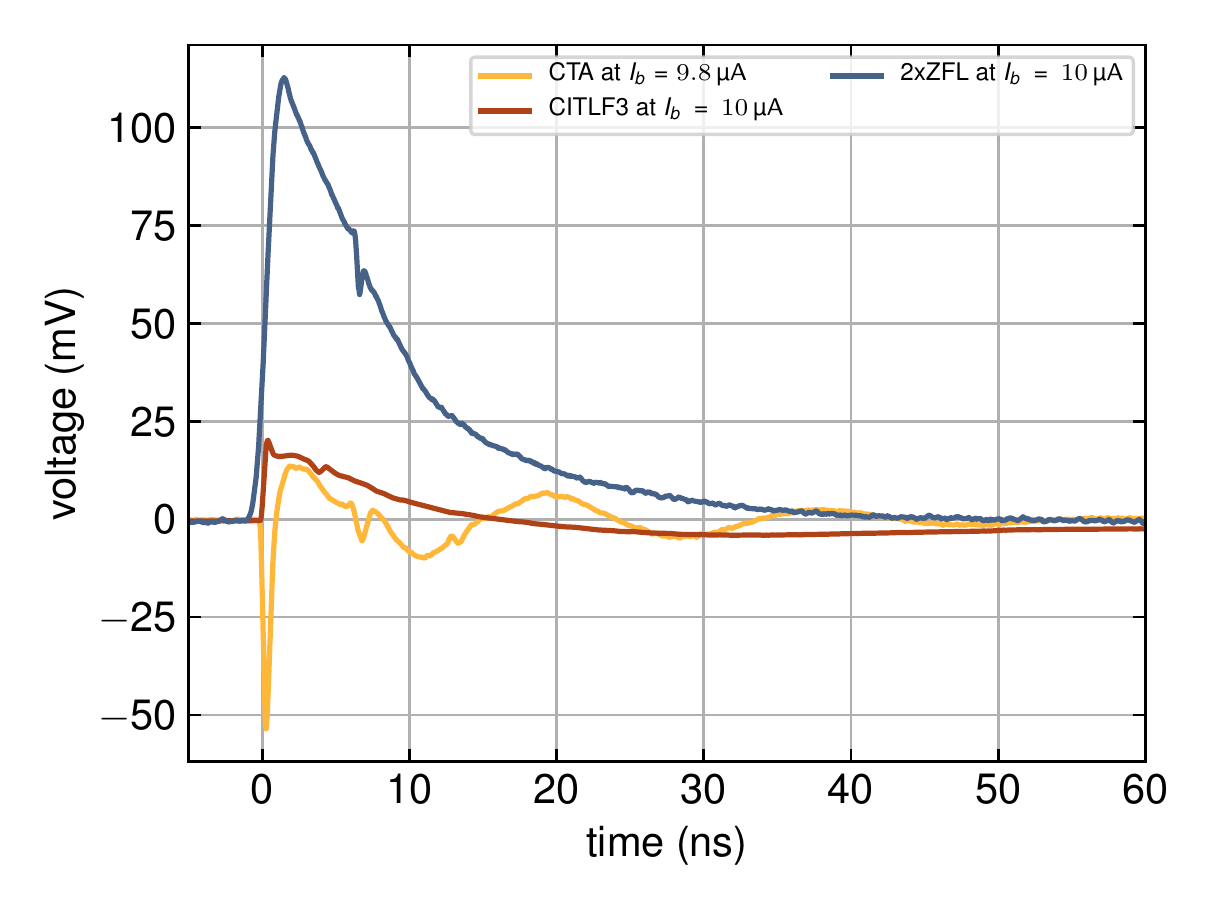}
\caption{\label{fig:click_shape_all} Voltage traces for different readout schemes recorded using a fast oscilloscope. The total count of amplification stages varies between the measurements, hence the first sharp signal edge used for timing may be rising or falling.}
\end{figure}

As all the electrical readout schemes discussed in Section D of the main text provide varying gain and bandwidth, the SNSPD response signal time trace shape varies with the readout scheme. In Figure \ref{fig:click_shape_all}, we plot the time traces of investigated readout schemes at an SNSPD bias current close to \SI{10}{\micro\ampere}. The readout comprising two Mini-Circuits ZFL-1000LN+ low noise amplifiers and a Mini-Circuits ZFBT-6GW-FT+ bias tee exhibits the highest gain and a low cut-off frequency of around 100\,kHz. The measured voltage trace clearly shows the typical exponential decay with no overshoot. The readout comprising a single ac-coupled CITFL3 (Cosmic Microwave Technology, Inc) and a Mini-Circuits ZFBT-6GW-FT+ provides significantly lower gain, and with a lower cut-off of around 10\,MHz the falling edge extends to the negative voltage regime. The scalable custom-made readout chain, consisting of a  cryogenic amplification stage with an L-R path to ground and a bias tee, and a room-temperature amplification stage, provides sufficiently high gain for the voltage signals to be read by state-of-the-art TCSPC units. The low cut-off frequency is around 50 MHz and hence the trailing exponential edge is visible as an oscillation after a first sharp rise. Note, that the envelope of this oscillation is still dominated by the time constant of the SNSPD, i.e. its kinetic inductance.

\section{Jitter histogram data}

\begin{figure}
\includegraphics[width=\columnwidth]{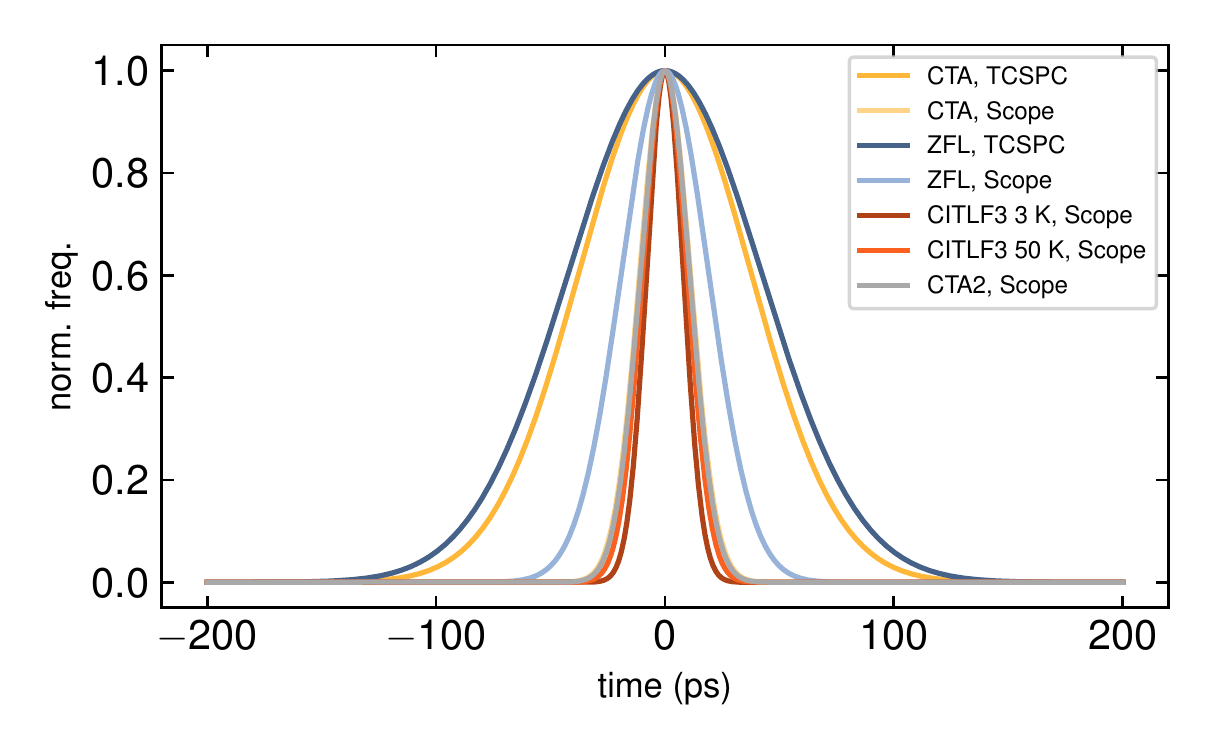}
\caption{\label{fig:jitter_histograms} Envelop of histograms of the lowest jitters recorded in the test system at 2.5\,K for different readout schemes. Note: Due to very similar timing uncertainty, the curves for the CTA (light yellow) and the CTA2 (grey) readout using the oscilloscope are overlapping.}
\end{figure}

The jitter values given in the main text are extracted from histogram data recorded using a high bandwidth Keysight Technologies Infiniium MSO-X 91304A mixed signal oscilloscope or a Picoquant MultiHarp 160 TCSPC unit. In Figure \ref{fig:jitter_histograms} the recorded histograms for the best jitter values from Figure 8 in the main text are given. 

\section{S21 parameter of the custom-made amplifiers}

\begin{figure}
\includegraphics[width=\columnwidth]{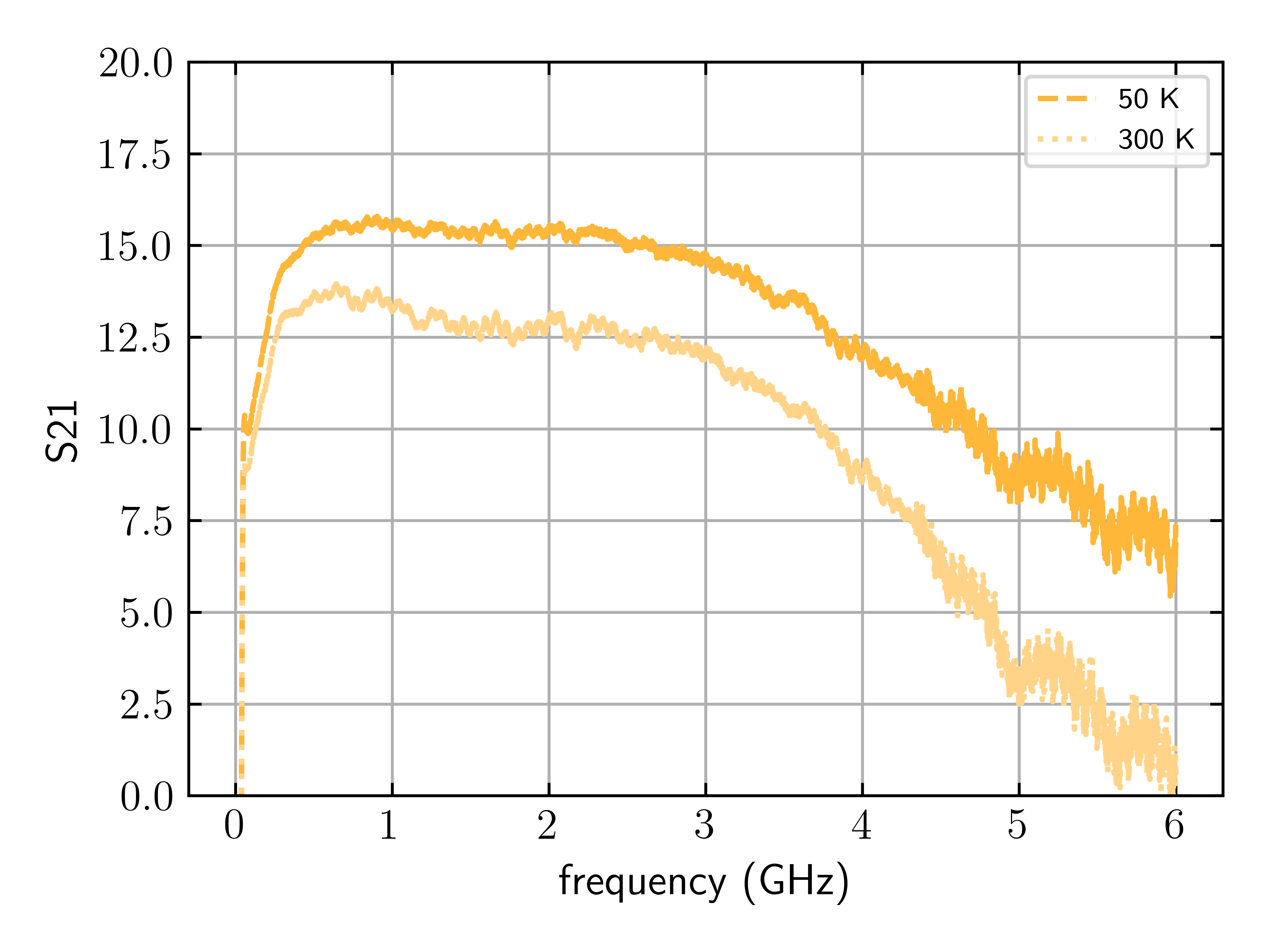}
\caption{\label{fig:amp2} S21 parameter of the single-stage custom-made cryogenic amplifier at room-temperature and at 50 K on the second stage of the closed-cycle cryostat.}
\end{figure}

\begin{figure}
\includegraphics[width=\columnwidth]{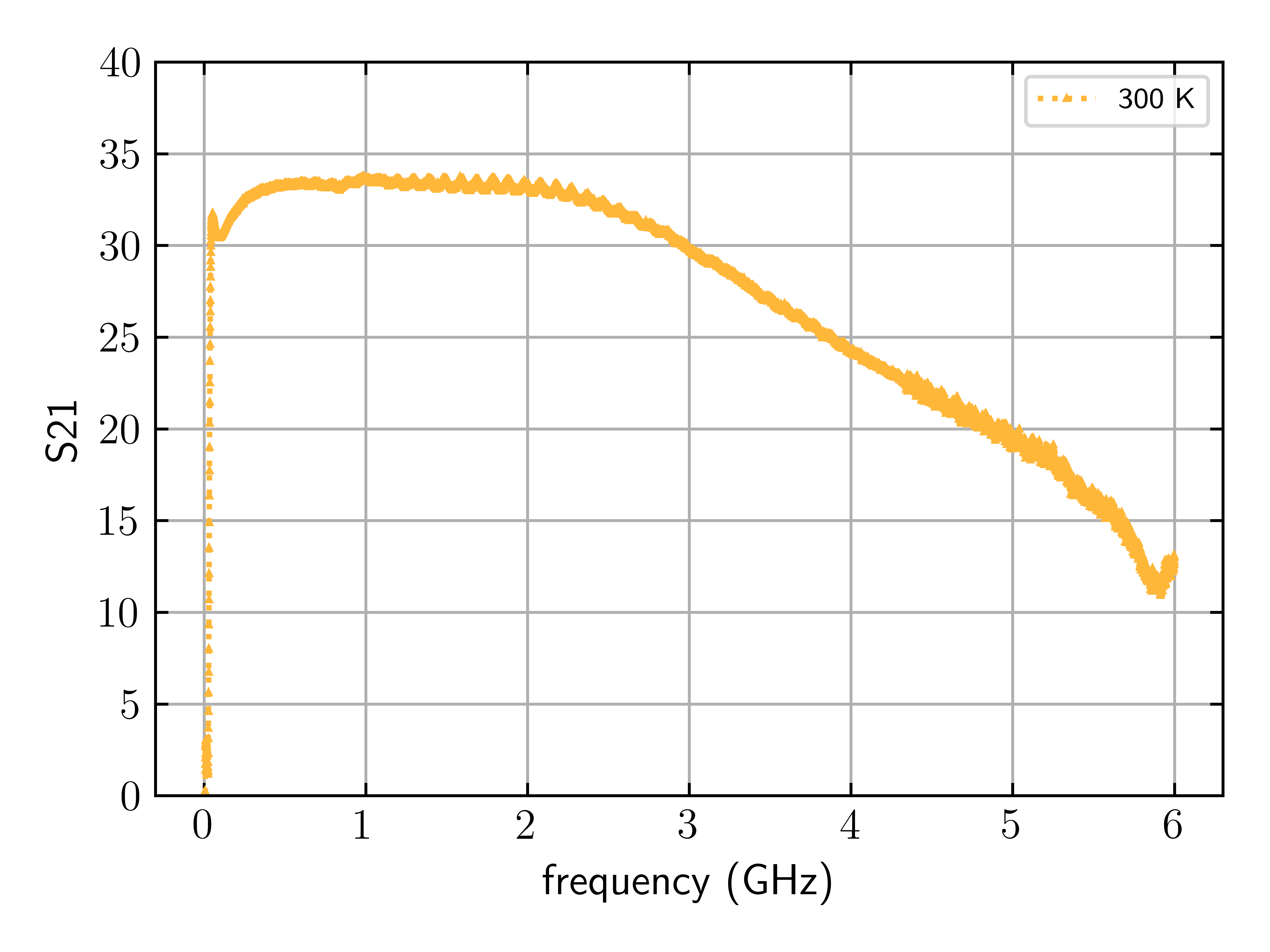}
\caption{\label{fig:amp4} S21 parameter of the custom-made room-temperature amplifier.}
\end{figure}

The S21 parameter of the single-stage custom-made cryogenic amplifier (CTA) is given in Figure \ref{fig:amp2}. A comparison between the operation at room-temperature and cryogenic temperatures at equal amplifier current of 3\,mA shows that the amplification is increased for cryogenic temperature operation. At cryogenic temperatures, the amplifier shows a 3\,dB bandwidth of more than 3\,GHz with a low cut-off around 200\,MHz and a high cut-off around 4\,GHz. We furthermore show the S21 parameter of the second amplification stage operated at room-temperature in Figure \ref{fig:amp4}. The room-temperature amplification stage exhibits a low cut-off of 200\,MHz and a high cut-off of around 3\,GHz.

\section{Jitter versus amplifier current}

\begin{figure}
\includegraphics[width=\columnwidth]{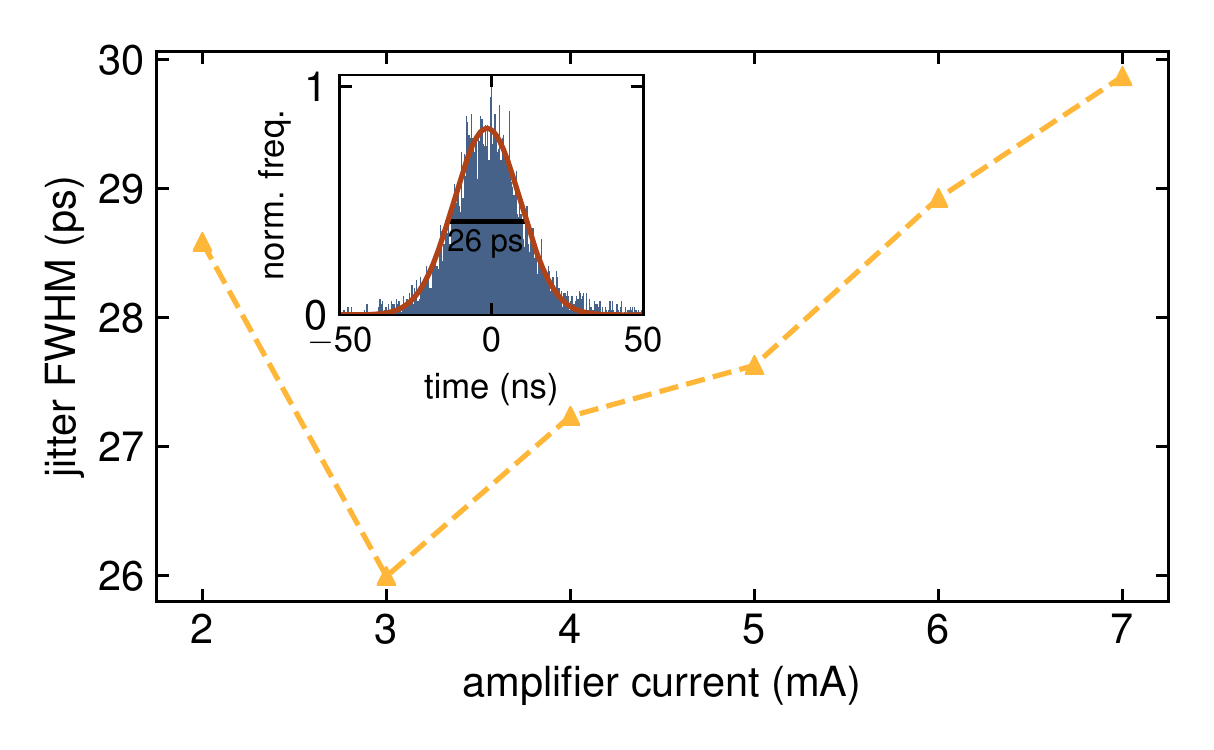}
\caption{\label{fig:jitter_versus_amp_current} Measured timing uncertainty with regard to the biasing conditions of the single-stage custom-made cryogenic amplifier. A minimum jitter is found at \SI{3}{\milli\ampere} bias current for the amplifier and a voltage of 1.8\,V. While the timing uncertainty is slightly higher at a current of \SI{2}{\milli\ampere}, lower power dissipation is beneficial for multi-channel systems. Inset: Histogram showing lowest jitter values measured using the single-stage custom-made amplifier with 26\,ps at \SI{3}{\milli\ampere} input current.}
\end{figure}

The jitter data for the single-stage custom-made cryogenic amplifier given in the main text is recorded for an amplifier current of \SI{3}{\milli\ampere}. By slightly reducing the amplifier current, an increase in the jitter is observed in exchange for a reduced power dissipation, which might be beneficial for systems equipped with a high number of amplifiers. Figure \ref{fig:jitter_versus_amp_current} shows the measured jitter versus the current. The minimal jitter at a current of \SI{3}{\milli\ampere} is measured at a total power dissipation of $\SI{1.8}{\volt}\cdot\SI{3}{\milli\ampere}=\SI{5.4}{\milli\watt}$.

\section{Multi-channel cryogenic amplifier PCB and cryostat integration}

\begin{figure}
\includegraphics[width=\columnwidth]{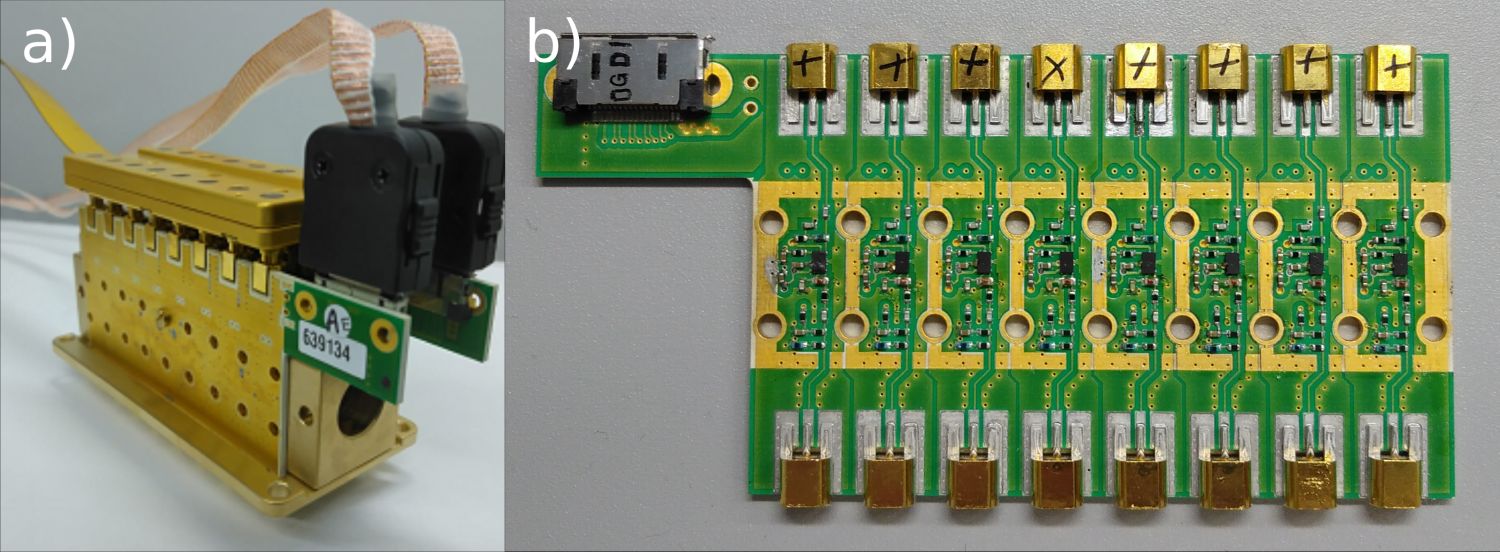}
\caption{\label{fig:cryo_amps} a) Picture of one 16-channel subassembly housing two 8-channel cryogenic preamplifier boards with attached 16-channel coaxial ribbon cables and multi-channel wooven-loom cables for DC supply. b)  Picture of a 8-channel cryogenic amplifier PCB.}
\end{figure}

The 64 cryogenic preamplification channels are split in four 16-channel subassemblies. Each of these subassemblies is composed of two 8-channel cryogenic preamplifier PCBs and a mechanical mount tailored to the space requirements of the cryostat (see Figure \ref{fig:cryo_amps}). The mount features chambers to provide sufficient RF-shielding between individual channels of one PCB and a simple mechanism to allow connecting and disconnecting the compact flex cables from the assembly without damage to the PCBs.

\section{Maximum count rate and latching}

\begin{figure}
\includegraphics[width=\columnwidth]{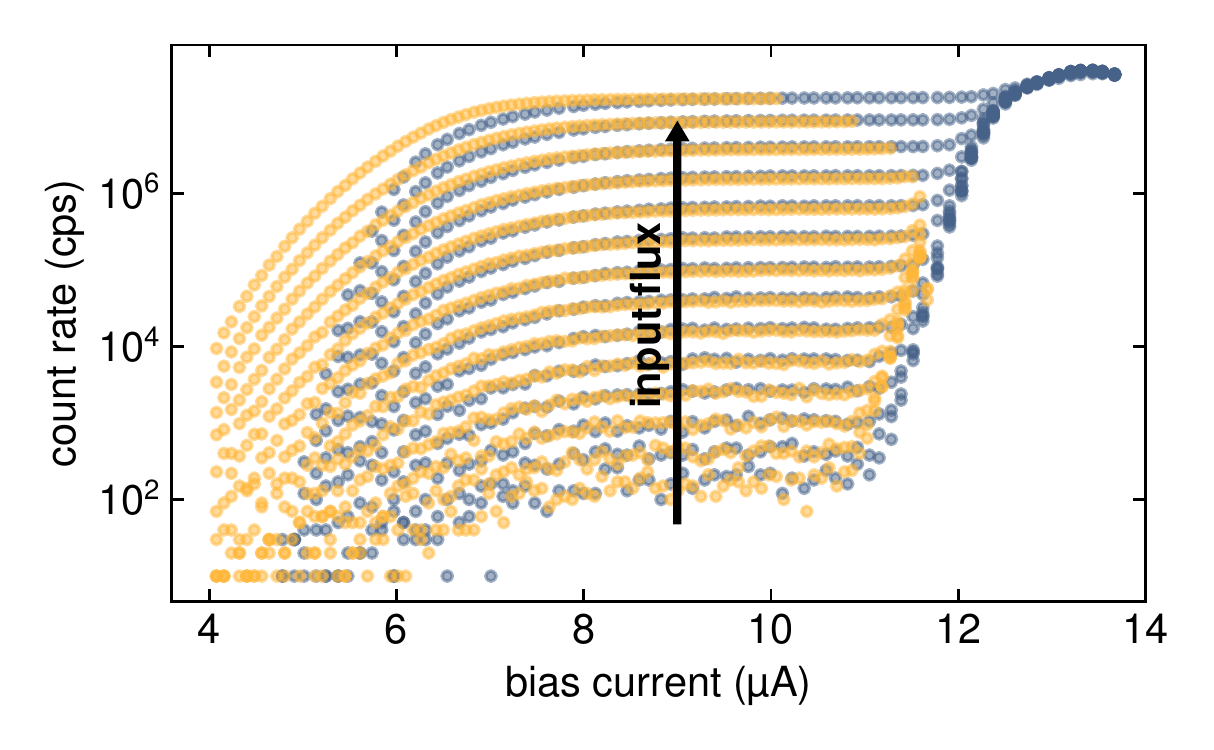}
\caption{\label{fig:countrate_all} Measured curves for the count rate versus SNSPD bias current for different input photon fluxes (+4\,dB input flux per measurement) for a readout scheme comprising commercially available readout electronics without L-R path to ground (yellow) and the scalable custom-made readout featuring an L-R path to ground (blue).}
\end{figure}

The maximum bias current of our custom-made readout circuit does not depend on the incident photon flux or the count rate of the detector due to the L-R path to ground. This allows to bias at high constant bias currents even under strong or strongly varying detector illumination. This has the advantage that the operating current does not need to be changed for different incident photon fluxes. Furthermore, the detector is also protected from latching due to voltage spikes in the biasing and readout circuitry resulting from non-optimal isolation, which allows for stable long-term operation beyond laboratory conditions. In Figure \ref{fig:countrate_all}, measured count rate versus bias current curves for different input photon fluxes and readout schemes are given. The data has been recorded with a single readout-channel and the SNSPD operating at 2.5\,K. Figure 12 of the main text shows selected data at different bias conditions (\SI{8}{\micro\ampere}, \SI{9}{\micro\ampere} and \SI{10}{\micro\ampere}).

\end{document}